\documentstyle[12pt,epsfig]{article}

\columnwidth\textwidth

\newcommand{\bildchen}[3]{
\begin{center}                                        %
\begin{flushleft}                                     %
\makebox{\Large $\displaystyle #2$}                   %
\end{flushleft}                                       %
\mbox{{\epsfig{figure=#1,width=12.cm,
bbllx=1.8cm,bblly=9.2cm,bburx=20.cm,bbury=19.cm}}}    %
\end{center}                                          %
\begin{flushright}                                    %
   {\Large $\displaystyle #3$ \hspace*{5ex}}          %
\end{flushright}}                                     %
\newcommand{\bildchenp}[3]{
\begin{center}                                        %
\begin{flushleft}                                     %
\makebox{\Large $\displaystyle #2$}                   %
\end{flushleft}                                       %
\mbox{{\epsfig{figure=#1,width=10.cm,
bbllx=0pt,bblly=0pt,bburx=288pt,bbury=288pt}}}        %
\end{center}                                          %
\begin{flushright}                                    %
   {\Large $\displaystyle #3$ \hspace*{5ex}}          %
\end{flushright}}                                     %

\begin{document}
\pagenumbering{arabic}
\everymath={\displaystyle}
\vspace*{-2mm}
\thispagestyle{empty}
\hfill BUTP--96/9\\
\mbox{}
\hfill HUTP95/-A051\\
\mbox{}
\hfill TTP96--02\\
\mbox{}
\hfill hep-ph/9604279\\
\mbox{}
\hfill  April 1996  \\
\begin{center}
  \begin{Large}
  \begin{bf} {\Large \sc 
Tau Decays and Chiral Perturbation Theory}
   \\
  \end{bf}
  \end{Large}
  \vspace{0.8cm}
   Gilberto Colangelo\\[2mm]
   {\em Institut f\"ur Theoretische Physik\\
    Universit\"at Bern\\
    Sidlerstrasse 5\\
    CH-3012 Bern, Switzerland}\\[5mm]
   Markus Finkemeier \\[2mm]
   {\em Lyman Laboratory of Physics\\
        Harvard University\\
        Cambridge, MA 02138, USA}\\[5mm]
   Res Urech\\[2mm]
   {\em Institut f\"ur Theoretische Teilchenphysik\\
        Universit\"at Karlsruhe\\
        D-76128 Karlsruhe, Germany}\\[5mm]
{\bf Abstract} 
\end{center}
\noindent
In a small window of phase space, chiral perturbation theory
can be used to make standard model predictions for tau decays
into two and three pions.
For $\tau\to2\pi\nu_\tau$, we give the analytical result for 
the relevant form factor $F_V$ up to two loops,
then calculate the differential spectrum and compare with available
data.
For $\tau\to 3 \pi \nu_\tau$, we have calculated the hadronic matrix 
element to one loop.
We discuss the decomposition of the three pion states into partition 
states and
we give detailed predictions for the decay in terms of  
structure functions.
We also compare with low energy predictions of meson dominance models.
Overall, we find good agreement, but also
some interesting discrepancies, which  
might have consequences beyond the limit of validity
of chiral perturbation theory.
\newpage

\section{Introduction}       
Semileptonic decays of the heavy tau lepton into a tau neutrino and
a hadronic system offer a unique laboratory to study the
standard model and especially low energy QCD.
A variety of multi-meson final states with invariant masses 
from the production threshold up to the tau mass of about $1.8 \,
\mbox{GeV}$ 
can be studied.
These final states can also be produced using hadronic initial states 
(such as pion-nucleon or nucleon-nucleon collisions).
The production in tau 
decays, however, is advantageous in 
that the initial state is simple, clean and well understood.
Whereas some of the final states (e.g. two or four pions) can also be 
produced using an electromagnetic current, i.e. electron-positron 
annihilation, other states (e.g. three pions in an $I=1$ state) can only
be produced through the weak current.
And in the case of the states which can be produced both in tau decays
and in electron-positron annihilation, such as the two and four pion 
final states, 
tau days now compete well in statistics with
electromagnetic production. 

For inclusive semileptonic tau decays, the tau mass of about
$1.8\,\mbox{GeV}$ 
is large enough to allow the application of perturbative QCD 
and in fact offers a unique possibility to measure the strong coupling
constant $\alpha_s(\mu)$ at the low scale $\mu = M_\tau$ \cite{tauQCD}.
In the case of exclusive semileptonic decays, calculations based on a
systematic use of the QCD Lagrangian are not available up to
now. 
In these decays, the probe testing the hadronic current carries a
momentum $Q$ which is mainly in the intermediate energy region, the most
difficult one for the study of strong interactions. 
In this region theoretical predictions have been obtained by using some
kind of phenomenological models or approximate methods, such as
quark models \cite{Isg,Iva}, vector meson dominance
\cite{fischer80,Kue90,Gom90,Dec92,Dec94,vmd}, tree level calculations from
effective Lagrangians \cite{braaten}, or unitarization of current
algebra results \cite{truong}.

A small fraction of these decays, however, happens with very low
$Q^2$, below the mass of the lightest resonance. In this region the
only active fields are the pseudoscalar mesons and one can use an
effective Lagrangian to describe their interactions. This effective
theory, called Chiral Perturbation Theory (CHPT), is a systematic
method to calculate QCD matrix elements at low energy by means of an
expansion in powers of the external hadronic momenta 
\cite{weinberg79,gasser84,reviews}.

The limits of applicability of CHPT do not allow to give predictions for 
integrated decay rates, which would involve 
$Q^2$ up to $M_\tau^2$. 
Furthermore, tau decays into more than a single pion are dominated by 
resonant intermediate states, such as the $\rho(770)$ in the
two pion channel and the $a_1(1260)$ in the three pion channel.
And so there are only relatively few events with small $Q^2$. 
For these reasons, in the past, CHPT has been considered as not very
interesting for tau decays. 

However, with the present high luminosity machines such as LEP and CESR,
and even more so with future b and perhaps tau-charm factories,
tau physics has turned into an era of precision measurements, exploring
very small branching ratios and studying details of differential
distributions. Thus the small $Q^2$ regime which is interesting for
CHPT is now becoming accessible.

We would like to mention that there is another small corner of phase
space where a different
systematic expansion of the hadronic current becomes possible.
Heavy meson chiral perturbation
theory \cite{hmcpt} can be applied to tau decays 
into a vector meson and a pion (such as
$\tau\to\rho\pi\nu_\tau$, $\tau\to K^* \pi\nu_\tau$), 
if the momentum of the pion is small in the vector meson rest frame 
\cite{wise}.
A complete calculation which includes vector meson decay and interference
effects between different vector meson amplitudes, however, is still 
missing in this approach.

Another reason why CHPT is relevant to tau decays is 
the fact that it can be
used to test phenomenological models or fix some of their parameters.
Indeed, the $O(p^2)$ prediction of CHPT in 
the limit of vanishing quark masses has been used to normalize vector meson
dominance models in \cite{fischer80,Kue90,Dec92,Dec94,vmd}.
In the present paper we will extend the CHPT prediction to higher order
in $p^2$, and it will be a severe test for models if they correctly
reproduce these higher orders.

The expansion parameter of CHPT is $Q^2/ (4 \pi
F_\pi)^2$, with $4 \pi F_\pi = 1.2  \,\mbox{GeV}$, so we are interested
in $\sqrt{Q^2}$ below $500 \dots 600 \,\mbox{MeV}$. 
Hadronic final states with a single pion or kaon can be predicted
directly from $F_\pi$ and $F_K$ \cite{Tsai}, and so there is nothing
interesting CHPT could teach us here. Final states with two and three
pions allow for a reasonably large region of $Q^2$ between threshold
and the limit of applicability of CHPT. Already with four pions this region
has almost disappeared.
Moreover, the phase space for a $n$ pion hadronic state, with $Q^2$
close to threshold $Q^2 \to (n M_\pi)^2$, opens proportional to
\begin{equation}
   \Big( \sqrt{Q^2} - n M_\pi \Big)^{(3 n - 5)/2}
\end{equation}
(see Sec.~3.1 below).
The exponent is $1/2$ for two pions, $2$ for three pions and $7/2$ for
four pions. So in the case of four pions the small interesting region
for CHPT is even more suppressed by the phase space.
As for final states with kaons, the threshold for a pion-kaon state is
$M_\pi + M_K = 634\,\mbox{MeV}$ . 
Furthermore the $K^*(892)$ resonance is very close \cite{fm95}.

Therefore we find reasonable to try a CHPT calculation only for 
the $2\pi$ and $3\pi$ final states
and these are the ones we will discuss in this paper.
The two pion state is determined by only one form factor, namely the
vector pion form factor. This can be measured in a few other
processes, like $e^+e^- \to 2\pi$ or $ \pi e $ scattering. Tau decays
can provide an interesting cross check measurement, and we will
investigate whether tau decays may become competitive in statistics.
The three pion state, however, can only be produced in tau decay,
and has a very rich and interesting substructure, which we will
study in detail. 

Our paper is organized as follows:
A brief review of chiral perturbation theory is given in Sec.~2.
In Sec.~3, we discuss the general structure of the phase space, the
hadronic matrix element for two and three pions, and the differential
decay rate. In this section we also discuss  general properties 
of the three pion final state, regarding isospin invariance,
the classification in terms of partition states and the definition 
of structure functions.
We then calculate the hadronic matrix elements
in CHPT in Sec.~4. Sec.~5 is dedicated to our numerical results, and
in Sec.~6 we state our conclusions.
In the Appendix we summarize our main conventions,
display the decomposition of the three pion form factors into partition 
states and collect the main formulae regarding the definition of the 
structure functions.
%
%
\section{Chiral Perturbation Theory} 
\label{CHPT}

The relevant matrix elements for $\tau$ decays into pions are of the
form:
\begin{equation}
\langle \pi^{i_1}(p_1) \ldots \pi^{i_n}(p_n) {\rm out}|I^k_\mu(0)|0 \rangle
\end{equation}
where $I^k_\mu=V^k_\mu=\frac{1}{2} \bar{q} \tau^k \gamma_\mu q$ when
$n$ is even, and $I^k_\mu=A^k_\mu=\frac{1}{2} \bar{q} \tau^k \gamma_\mu
\gamma_5  q$ when $n$ is odd. A by now standard method to calculate
such matrix elements in QCD at low energy is 
Chiral Perturbation Theory, CHPT.
In this framework
one uses an effective Lagrangian that respects the chiral symmetry
properties of QCD, and that has the pions as the only ``active''
fields. Of course this Lagrangian is expected to be valid only up to
energies which are well below the threshold for the production of
heavier hadronic states. For more details about this method we refer
the reader to the fundamental paper by Gasser and Leutwyler
\cite{gasser84} and to a number of excellent reviews which are
presently available in the literature \cite{reviews}. Here we simply
sketch the basic ideas and introduce the relevant notation.

We consider the effective Lagrangian relative to two flavors in the
isospin limit  $\hat{m} = m_u = m_d$. This Lagrangian contains an infinite
number of terms; however it can be expanded in powers of derivatives
and quark masses. One power of the quark mass will be counted as two
powers of derivatives\footnote{This means we are applying the
``standard'' CHPT counting rule. For a different counting rule,
leading to a different ordering in the effective Lagrangian 
   (the so-called ``generalized'' CHPT), see Ref.
   \cite{GCHPT}.}.
One will then have:
\begin{equation}
{\cal L}_{\rm eff} = {\cal L}_2+ \hbar {\cal L}_4+\hbar^2{\cal L}_6+\ldots
\; . 
\end{equation}
The leading order Lagrangian starts at $O(p^2)$ and is the nonlinear
$\sigma$-model Lagrangian in the presence of external fields, which we
represent here in matrix form,
$a_\mu=a^k_\mu \tau^k$, $v_\mu=v^k_\mu \tau^k$:
\begin{eqnarray}
\label{efflagr}
{\cal L}_2 &=& \frac{F^2}{4}\langle D^\mu U D_\mu U^\dagger 
           + M^2 \left(U  + U^\dagger \right) \rangle \nonumber \\[4mm]
D_\mu U &=& \partial_\mu U - i (v_\mu + a_\mu) U + i U (v_\mu -
a_\mu)\nonumber \\ 
M_\pi^2 &=& M^2 \left[1 + O(\hat{m})\right]\nonumber \\
M^2 &=& 2B\hat{m} \nonumber \\
F_\pi &=& F \left[1 + O(\hat{m})\right]
\end{eqnarray}
$B$ is proportional to the quark condensate $\langle 0| \bar{u}u |0
\rangle$ and the unitary $2 \times 2$ matrix $U$ contains the pion fields, 
\begin{eqnarray}
U &=& \sigma + i \frac{\phi}{F}
\hspace{2cm}\sigma^2 + \frac{\phi^2}{F^2} = {\bf 1} \nonumber \\
\phi&=&\left(
\begin{array}{cc}
\pi^0           & \sqrt{2}\pi^{+} \\
\sqrt{2}\pi^{-} & -\pi^0
\end{array}
\right) \; \; .
\end{eqnarray}
The external fields $v^k_\mu$ and $a^k_\mu$, which we
have introduced, have to be treated as external sources for the {\it
quark} currents $V^k_\mu$ and $A^k_\mu$ respectively; in other words,
the currents coupled to $v^k_\mu$ and $a^k_\mu$ in the effective
Lagrangian are the low energy representation of the quark currents. In
this framework these currents are expanded in powers of derivatives
and quark masses, and are nonlinear in the pion fields, e.g. the
axial vector current, 
\begin{equation}
A_\mu^k = \frac {iF^2}{4} \langle \tau^k (U^\dagger D_\mu U -
U D_\mu U^\dagger) \rangle + O(p^3) = 
\left[ -F \partial_\mu \phi^k+O(\phi^3) \right] + O(p^3) \; \; .
\end{equation}

Despite the fact that this Lagrangian is nonrenormalizable, one can
use it to calculate matrix elements with the standard perturbation
theory. As we have emphasized in Eq.~(\ref{efflagr}) the expansion
parameter is $\hbar$. This automatically produces an expansion of
the matrix elements in powers of momenta and quark masses.
As for the matrix elements in question, tree diagrams from ${\cal
L}_2$ generate leading order contributions
\cite{fischer80,Kue90,Dec92,Dec94}, 
while one-loop diagrams yield terms at next-to-leading order. The
occurring divergences in the loop contributions (in $d=4$ dimensions)
can be absorbed by introducing the effective Lagrangian at $O(p^4)$
\cite{gasser84}, 
\begin{eqnarray}
{\cal L}_4 &=& \frac{1}{4} l_1 \langle D^\mu U D_\mu U^\dagger \rangle^2 
             + \frac{1}{4} l_2 \langle D_\mu U D_\nu U^\dagger \rangle
                           \langle D^\mu U D^\nu U^\dagger \rangle
\nonumber \\ 
           &&+ \frac{1}{16} l_3 M^4 \langle U + U^\dagger\rangle^2 
             + \frac{i}{2} l_4 M^2 \langle a_\mu \left(D^\mu U  
                           -  D^\mu U^\dagger \right) \rangle \nonumber \\
           &&+ l_5 \langle F_R^{\mu\nu} U F_{L\;\mu\nu} U^\dagger \rangle
                + \frac{i}{2} l_6\langle F_R^{\mu\nu} 
                                       D_\mu U D_\nu U^\dagger 
                + F_L^{\mu\nu} D_\mu U^\dagger D_\nu U \rangle 
			+ \ldots \nonumber \\[4mm]
F_{R,L}^{\mu\nu} &=& \partial^\mu (v^\nu \pm a^\nu) 
                     - \partial^\nu (v^\mu \pm a^\mu) 
                     - i [v^\mu \pm a^\mu,v^\mu \pm a^\mu]
\end{eqnarray}
where we omitted terms which contain external fields only.  
Since we disregard singlet vector and axial vector currents
(i.e. $\langle v_\mu \rangle = \langle a_\mu \rangle = 0$) there is no
contribution from the anomaly at $O(p^4)$ \cite{gasser84}.

The coupling constants $l_i$ are split in a divergent and a finite piece
and are scale-independent by definition,
\begin{eqnarray}
l_i &=& \gamma_i \lambda + l_i^r(\mu), \hspace{2cm}
\lambda = \frac{\mu^{d-4}}{16\pi^2} \left\{\frac{1}{d-4}
                   -\frac{1}{2}[\ln 4\pi+\Gamma'(1) +1 ] \right\} 
                   \nonumber \\[2mm]
\mu\frac{d}{d\mu}l_i &=& \gamma_i \frac{\mu^{d-4}}{16\pi^2}
                         + \mu\frac{d}{d\mu} l_i^r(\mu) 
                         + O(d-4) = 0
\end{eqnarray}
The finite parts have been determined phenomenologically for the first
time by Gasser and
Leutwyler \cite{gasser84}. We adopt their notation and use instead of
the $ l_i^r(\mu)$, the finite and scale-independent quantities
$\bar{l}_i$,
\begin{equation}
\bar{l}_i = \left(\frac{\gamma_i}{32\pi^2}\right)^{-1}
            l_i^r(\mu) \; - \; \ln \frac{M^2}{\mu^2}
\end{equation}
The up to date values for the relevant $\bar{l}_i$ and the
corresponding $\gamma_i$ are listed in Table \ref{li}. 
\begin{table}[t]
\begin{center}
\begin{tabular}{|c|r|c|}
\hline &&\\[-3mm]
$i$ & $ \hspace{0.5cm} \bar{l}_i$ \hspace{0.5cm} &  $\gamma_i$  \\[2mm] 
\hline &&\\[-3mm]
1   & $-1.7$ $\pm$ 1.0  &  1/3  \\[2mm]
2   &  6.1 $\pm$ 0.5     &  2/3  \\[2mm]
3   &  2.9 $\pm$ 2.4     &  $-1/2$ \\[2mm]
4   &  4.3 $\pm$ 0.9     &  2              \\[2mm]
6   & 16.5 $\pm$ 1.1     &  $-1/3$ \\[2mm]
\hline
\end{tabular}
\end{center}
\caption[]{The set of $\bar{l}_i$ and corresponding $\gamma_i$ which we
need for the calculation of the matrix elements in question. The
values are taken from \cite{bcg} for $i=1,2$, and from \cite{gasser84}
for all the others. 
The $\gamma_i$ determine the relation between the $\bar{l_i}$ and the
$l^r_i(\mu)$.} 
\label{li}
\end{table}
For completeness we give the expressions for the pion decay constant and
the pion mass up to and including $O(p^4)$,
\begin{eqnarray}
F_\pi &=& F \left[ 1 + \frac{M^2}{16\pi^2 F^2}\bar{l}_4 
                     + O(M^4) \right] \nonumber \\[2mm]
M_\pi^2 &=& M^2 \left[ 1 - \frac{M^2}{32\pi^2 F^2}\bar{l}_3 
                         + O(M^4) \right]
\end{eqnarray}
The mass splitting $M_{\pi^\pm}^2 - M_{\pi^0}^2$ is proportional to  $(m_u
- m_d)^2$ and thus may be neglected.
In the numerical evaluation, we will use $M_\pi = 139.57 \, \mbox{MeV}$
and%
  \footnote{This means that we are neglecting $O(\alpha)$ corrections
to the decay width $\Gamma(\pi\to\mu\nu_\mu)$ from which
$F_\pi$ is extracted, for more details see Ref. \cite{fpi}.}
$F_\pi = 93.1\,\mbox{MeV}$.

If one wants to go beyond the next-to-leading order, one has to
calculate two loop diagrams with ${\cal L}_2$, and one loop diagrams
with one vertex from ${\cal L}_4$. Again these diagrams will be
divergent, but this is not a problem since at the same order one
has contributions from tree diagrams with the ${\cal L}_6$
Lagrangian (which has been constructed  in the case of three lights
flavors 
\cite{fearing96}). By defining appropriately 
the new coupling constants occurring in this Lagrangian, one is able to
remove the divergences at the next--to--next--to--leading order, and get
finite matrix elements. In order to have numerical predictions at this
level one has to find a way to pin down or at least to estimate the
finite parts of the new low energy constants. At the moment this has
not been done yet in a systematic way: in Sec. \ref{twopion}, when
calculating the matrix element of the $\tau$ decay into two pions to
two loops, we will show how in a specific case one can try to
circumvent this problem. 

\section{Phase Space, Matrix Elements and Structure Functions}
\subsection{Phase Space Considerations}
The two-pion phase space is given by
\begin{equation}
   \Phi_{2 \pi} = \frac{M_\tau^2}{128 \pi^3} 
   \int \frac{dQ^2}{M_\tau^2} \left(1 - \frac{Q^2}{M_\tau^2} \right)
   \left( 1 - \frac{4 M_\pi^2}{Q^2} \right)^{1/2}
\end{equation}
(the reader is referred to App.~\ref{app1} for our conventions)
where the integral is from threshold $4 M_\pi^2$ up to $M_\tau^2$. 
Close to the two pion threshold, this implies
\begin{equation}
  \frac{d \Phi_{2 \pi}}{d Q^2} \to \frac{1}{128 \pi^3}
  \frac{1}{ M_\pi^{1/2}} 
  \left(1 - \frac{4 M_\pi^2}{M_\tau^2} \right)
  \Big(\sqrt{Q^2} - 2 M_\pi \Big)^{1/2}
\end{equation}
The three pion phase space is
\begin{eqnarray}
  \Phi_{3 \pi} & = &\frac{M_\tau^2}{2048 \pi^5} \int_{9
                     M_{\pi}^2}^{M_\tau^2} 
   \frac{dQ^2}{M_\tau^2} \left(1 - \frac{Q^2}{M_\tau^2} \right) 
  \frac{1}{Q^2} 
  \int_{4 M_\pi^2}^{(\sqrt{Q^2} - M_\pi)^2} \frac{ds_1}{s_1}
\nonumber \\ 
\nonumber \\ 
 & & \quad \times
  [ (s_1 - Q^2 - M_\pi^2)^2 - 4 Q^2 M_\pi^2]^{1/2}
  [s_1 (s_1 - 4 M_\pi^2)]^{1/2}
\end{eqnarray}
Close to threshold $Q^2 \to 9 M_\pi^2$ this can be approximated by
\begin{equation}
  \frac{d \Phi_{3 \pi}}{d Q^2} \to \frac{1}{2^{10} 3^{3/2} \pi^4}
    \left( 1 - \frac{9 M_\pi^2}{M_\tau^2} \right)
   \Big( \sqrt{Q^2} -3 M_\pi \Big)^2
\end{equation}
By induction we can show \cite{Byk} that 
the phase space for a $n$ pion hadronic state, with $Q^2$
close to threshold, $Q^2 \to (n M_\pi)^2$, opens proportional to:
\begin{equation}
  \frac{d \Phi_{n \pi}}{d Q^2} \propto 
   \Big( \sqrt{Q^2} - n M_\pi \Big)^{(3 n - 5)/2}
\end{equation}
For $n=2$, the exponent is $1/2$, for $n=3$ it is $2$, recovering
the above results.

It is clear that in general, the more pions there are, the slower
the phase space opens at threshold.
Therefore for four and more pions, it is not only the high invariant
hadronic mass, but also the behavior of the phase space at threshold
which prevents the application of CHPT.

\subsection{Two Pion Differential Decay Rate}
The hadronic matrix element of the decay into two pions is characterized
by a single form factor, $F_V(Q^2)$,
\begin{equation}
  H^\mu = \langle \pi^-(p_1) \pi^0(p_2) | V^\mu - A^\mu | 0 \rangle 
  = \sqrt{2} (p_1 - p_2)^\mu F_V\left(Q^2\right)
\end{equation}
Only $|F_V(Q^2)|$ can be measured, and it can be obtained by 
measuring the differential distribution in $Q^2$ using
\begin{equation}
  \frac{d \Gamma_{2 \pi}}{\Gamma_e}(Q^2)  =
  \frac{\cos^2 \theta_c}{2}
  \frac{dQ^2}{M_\tau^2} 
  \left( 1 - \frac{Q^2}{M_\tau^2} \right)^2
  \left(1 + \frac{2 Q^2}{M_\tau^2} \right)
  \left( 1 - \frac{4 M_\pi^2}{Q^2} \right)^{3/2} 
  | F_V(Q^2) |^2
\end{equation}
Here we have normalized to the electronic branching ratio of the tau:
\begin{equation}
   \Gamma_e = \frac{G_F^2 M_\tau^5}{192 \pi^3}
\end{equation}
%
\subsection{Structure Functions and the Three Pion Differential
Decay Rate}
\subsubsection{Form Factors and Isospin Relations}
\label{FF}
The most general form of the hadronic matrix elements for the tau
decays into the $2 \pi^- \pi^+$ and $2 \pi^0 \pi^-$ final states, 
compatible with the requirements of Lorentz, isospin and $G$ parity
invariance and Bose symmetry is given in terms of three functions $F,
\; G$ and $H$ which have to satisfy the following property:
\begin{eqnarray}\label{symm}
   F(s_2,s_1,s_3) & = & + F(s_1,s_2,s_3)
\nonumber \\ 
   G(s_2,s_1,s_3) & = & + G(s_1,s_2,s_3)
\nonumber \\ 
   H(s_2,s_1,s_3) & = & - H(s_1,s_2,s_3) \; .
\end{eqnarray}

The matrix elements are then given by
\begin{eqnarray}\label{decomp}
\lefteqn{\langle \pi^0 (p_1) \pi^0 (p_2) \pi^- (p_3)|
A_\mu^- (0)|0\rangle } \nonumber \\[2mm]
        & = & G(s_1,s_2,s_3) (p_1 + p_2)_\mu  
          + H(s_1,s_2,s_3) (p_1 - p_2)_\mu  
          + F(s_1,s_2,s_3) p_{3\,\mu} 
\\[4mm]
\lefteqn{\langle \pi^- (p_1) \pi^- (p_2) \pi^+ (p_3)|
A_\mu^- (0)|0\rangle } \nonumber \\[2mm]
        &=& G^{(+)}(s_1,s_2,s_3) (p_1 + p_2)_\mu  
          + H^{(+)}(s_1,s_2,s_3) (p_1 - p_2)_\mu  
          + F^{(+)}(s_1,s_2,s_3) p_{3\,\mu}  \nonumber \; .
\end{eqnarray}
The form factors for $2\pi^-\pi^+$ and $2\pi^0\pi^-$ are related
by isospin symmetry
\begin{eqnarray}\label{isospin}
F^{(+)}(s_1,s_2,s_3) &=& \left[ G(s_2,s_3,s_1) + G(s_3,s_1,s_2)
                    - H(s_2,s_3,s_1) + H(s_3,s_1,s_2)\right] \nonumber
\\[4mm]  
G^{(+)}(s_1,s_2,s_3) &=&   
           \frac{1}{2}\left[ F(s_2,s_3,s_1) + F(s_3,s_1,s_2)
           + G(s_2,s_3,s_1) + G(s_3,s_1,s_2) \right. \nonumber \\ 
&& \left.\hspace{1cm} + H(s_2,s_3,s_1) - H(s_3,s_1,s_2)\right] \nonumber
\\[4mm] 
H^{(+)}(s_1,s_2,s_3) &=& 
           \frac{1}{2}\left[ F(s_2,s_3,s_1) - F(s_3,s_1,s_2) 
           - G(s_2,s_3,s_1) + G(s_3,s_1,s_2) \right. \nonumber \\ 
        && \left.\hspace{1cm}  - H(s_2,s_3,s_1) 
                               - H(s_3,s_1,s_2)\right]
\end{eqnarray}

Alternatively one can use a decomposition of the matrix element into
only two functions, $F_1(s_1,s_2,s_3)$ and $F_S(s_1,s_2,s_3)$, with
$F_S$ even under the exchange of the first two arguments, and $F_1$ of
mixed behavior:
\begin{eqnarray}\label{decomp2}
\lefteqn{\langle \pi^0 (p_1) \pi^0 (p_2) \pi^- (p_3)|
A_\mu^- (0)|0\rangle } \nonumber \\[2mm]
&=& [F_1(s_1,s_2,s_3) (p_1 - p_3)^\nu
  +  F_1(s_2,s_1,s_3) (p_2 - p_3)^\nu ] T_{\mu \nu} + F_S Q_\mu
\nonumber \\[4mm]
\lefteqn{\langle \pi^- (p_1) \pi^- (p_2) \pi^+ (p_3)|
A_\mu^- (0)|0\rangle } \nonumber \\[2mm]
&=& [F_1^{(+)}(s_1,s_2,s_3) (p_1 - p_3)^\nu
  +  F_1^{(+)}(s_2,s_1,s_3) (p_2 - p_3)^\nu ] T_{\mu \nu} + F_S^{(+)}
Q_\mu \; ,
\end{eqnarray}
where
\begin{eqnarray}
T_{\mu\nu}   &=& g_{\mu\nu} - \frac{Q_\mu Q_\nu}{Q^2} \nonumber \\
Q_\mu        &=& (p_1 + p_2 + p_3)_\mu    \nonumber \\
Q^2          &=& s_1 + s_2 + s_3 - 3M_\pi^2 \label{q2} \; .
\end{eqnarray}
The decomposition into 
$F_1$ and $F_S$ has the advantage that these form factors correspond to
a definite overall spin (viz. $F_1$ corresponds to spin 1 and $F_S$  
to spin 0), and therefore the
structure functions (see next Sec. \ref{threepi}) are usually
expressed through $F_1$ and $F_S$. If $F$, $G$ and $H$ are known, we
can calculate $F_1$, $F_S$  through
\begin{eqnarray}
F_1(Q^2,s_1,s_2) &=& \frac{- F(s_1,s_2,s_3) + G(s_1,s_2,s_3)}{3}
                                 + H(s_1,s_2,s_3) \nonumber \\[2mm]
F_S(Q^2,s_1,s_2) &=& \alpha F(s_1,s_2,s_3) + (1 - \alpha) G(s_1,s_2,s_3)
                     - \beta H(s_1,s_2,s_3) \nonumber\\[4mm]
\alpha &=& \frac{s_1 + s_2 - 2M_\pi^2}{2Q^2} \nonumber \\[2mm]
\beta &=& \frac{s_1 - s_2}{2Q^2} \; .
\end{eqnarray}
A completely analogous relation holds for the form factors of the all
charged matrix element.

Let us emphasize two facts regarding the two charge modes.
Firstly, the two matrix elements for the final
states $2 \pi^- \pi^+$ and $2 \pi^0 \pi^-$ are not independent. If one knows
the matrix element for one of the two states, the other one can be calculated
using isospin symmetry.
Secondly, however, isospin symmetry does {\em not} require that the form
factors and decay rates are equal for the two modes. This fact can be seen
if we decompose the three pion states in terms of partitions \cite{Pai60}.
%
\subsubsection{Classification in Terms of Partitions}
\label{secpartitions}
In Ref. \cite{Pai60}, Pais introduced a classification of $N$ pion states
with overall isospin $I=0$ or $1$ in terms of correlation quantum numbers
$[N_1 N_2 N_3]$. The three integer quantum numbers $N_i$ are
partitions of the total number of pions $N$
\begin{eqnarray} &&
   N_1 \geq N_2 \geq N_3 \geq 0
\nonumber \\ &&
   N_1 + N_2 + N_3 = N
\end{eqnarray}
Each state $[N_1 N_2 N_3]$ is characterized by its symmetry property
under the exchange of some of the momenta $p_1, \ldots , p_N$. Such a
state is easily constructed with the help of a Young tableau: each Young
diagram must have three rows with $N_1, \; N_2$ and $N_3$ cells in the
first, second and third row, respectively. The cells
must then be filled with numbers going from 1 to $N$, with the only
rule that all the numbers in the rows (columns) must be organized in
increasing order from left to right (top to bottom). The rule to
construct a pion state from a tableau is very simple: one has to
symmetrize with respect to the exchange of the momenta with the
indices which are in a row, and antisymmetrize with respect to the
momenta with the indices which are in a column. The order of these
operations of symmetrization or antisymmetrization is not important,
but must be fixed once and for all.

Remarkably, all the states belonging to the same class defined by the
partition $[N_1 N_2 N_3]$ share some common properties about isospin
and charge distributions:
\begin{enumerate}
\item
the overall isospin $I$ is uniquely determined and it is $I=0$ if
$N_1-N_3$ and $N_2-N_3$ are both even, and $I=1$ otherwise;
\item
the states in a class $[N_1 N_2 N_3]$ are composed by $N_3$ subsystems
of three pions with $I=0$ and $N_2-N_3$ subsystems of two pions with
$I=1$, and $N_1-N_2$ remaining single pions (trivially $3 N_3
+2(N_2-N_3) + (N_1-N_2) = N$);
\item
the $N$ pion states which we are describing contain all possible charge
distributions (e.g. for $N=2$ and zero total charge, they would
contain both $\pi^+ \pi^-$ and $\pi^0 \pi^0$ states). The probability
of a state to contain a given charge distribution is a ``class
property", i.e. is uniquely determined by the partition $[N_1 N_2
N_3]$ to which it belongs.
\end{enumerate}

For a more detailed account of the properties of these $N$ pion
states we refer the reader to the original article by Pais \cite{Pai60}. 
We now
concentrate on the case of our interest $N=3$.

In this case we have three possible partitions:
$[300]$, $[210]$ and $[111]$. 
The $[111]$ corresponds to $\pi^+ \pi^0 \pi^-$ in an overall $I=0$ state
(e.g. from the decay $\omega\to 3 \pi$).
The remaining two partitions $[210]$ and $[300]$ have $I=1$ and so they 
can occur in $\tau \to 3 \pi \nu_\tau$. 

These two partitions differ in
their branching ratios into the two charge distribution states.
The $[210]$ decays equally into $2 \pi^- \pi^+$ and $2 \pi^0 \pi^-$:
\begin{equation} \label{eqn30}
   [210]: \qquad \frac{\mbox{BR}(2 \pi^- \pi^+)}
   {\mbox{BR}(2 \pi^0 \pi^-)} = 1
\end{equation}
whereas the $[300]$ state prefers the all charged mode
\begin{equation} \label{eqn31}
   [300]: \qquad \frac{\mbox{BR}(2 \pi^- \pi^+)}
   {\mbox{BR}(2 \pi^0 \pi^-)} = 4
\end{equation}
(\ref{eqn30},\ref{eqn31}) immediately lead to the inequalities
obtained in Ref. \cite{Gil85}: 
\[
\frac{1}{5} \leq \frac{\mbox{BR}(2 \pi^0 \pi^-)}{
\mbox{BR}(\mbox{all}\; (3\pi)^-)} \leq \frac{1}{2} 
\; , \; \;  \qquad
\frac{1}{2} \leq \frac{
\mbox{BR}(2\pi^-\pi^+)}{ \mbox{BR}(\mbox{all}\; (3\pi)^-)} \leq \frac{4}{5}
\;.
\]

Experimentally, the branching ratios of the $\tau$ into the two states
are equal within the errors, so that certainly the $[210]$ strongly
dominates and a possible 
small admixture of the $[300]$ state (which, as we will show below, is
required by CHPT), has not yet been established.
Note that if the decay occurs only via a decay chain $\tau \to a_1 \nu_\tau$,
$a_1 \to \rho \pi$ and $\rho \to \pi \pi$, as in vector meson dominance
models, there is only the $[210]$ state (because of the $\rho$
resonance, there is one two-pion subsystem with $I=1$, i.e. $N_2-N_3 =
1$), and both decay charge modes are produced with equal rates.

If one has an analytic expression for the form factors in each
of the two charge modes, one can easily construct the matrix element
of a given partition state, by following the rules which we described
above. The complete decomposition of each of the form factors $F, \;G$ and
$H$ into the three partition states (two states for the partition
$[210]$ and one for $[300]$) is described in App. \ref{partition}. 
Here we give in
an obvious notation
the decomposition only of $F_1$, since it is the most important form
factor in the numerical analysis:
\begin{eqnarray}
\label{ffpart}
F_1 &=&  \left[F_1^{[300]} + F_1^{[210]} \right] \nonumber
\\
F_1^{(+)} &=& \left[ 2 F_1^{[300]} -  F_1^{[210]} \right]
\; ,
\end{eqnarray}
where with $F_1^{[210]}$ we have indicated the sum of the two states
belonging to the $[210]$ class. From this decomposition the branching
ratios given in Eqs. (\ref{eqn30},\ref{eqn31}) follow, and since
$F_1^{\mbox{\tiny{VMD}}}=-F_1^{(+) \; \mbox{\tiny{VMD}}}$, it is clear
that the $[300]$ partition is absent in the VMD model.  
\subsubsection{Structure Functions and Differential Decay Rate}
The differential decay rate for a general hadronic decay is determined
by 
\begin{equation} \label{cca}
   d \Gamma(\tau\to 3\pi\nu_\tau) =
   \frac{G_F^2}{4 M_\tau} \cos^2 \theta_C L_{\mu\nu} H^{\mu\nu}
   \,dPS^{(4)}
\end{equation}
where the hadronic and leptonic tensors are 
\begin{equation}
   L_{\mu\nu} := L_\mu (L_\nu)^\dagger \; , \; \; \;
   H_{\mu\nu} := H_\mu (H_\nu)^\dagger \; ,
\end{equation} where $H_\mu = \langle \mbox{hadronic final
state}|A_\mu^-(0)|0\rangle $.
The decay is most easily analyzed in the hadronic rest frame,
and we can write
\begin{equation}
    L_{\mu\nu} H^{\mu\nu} = \sum_{X} L_X W_X
\end{equation}
In general, $H^{\mu\nu}$ can be characterized by 16 independent real 
functions. In our case of a three pion final states, there are restrictions
due to $G$ parity and Bose symmetry, which leave 9 independent structure 
functions $W_X$. 
These hadronic structure functions $W_X$ depend on the kinematics
only through the hadronic invariants $s_1$, $s_2$ and $Q^2$. 
The angular dependence is 
contained fully in the corresponding leptonic $L_X$. For details, see 
App.~\ref{app2} below and \cite{KM1,KM2}.

There are four structure functions, $W_A$, $W_C$, $W_D$, and $W_E$,
which arise from the
spin-1 part of the hadronic current, i.e. they depend on $F_1(Q^2,s_1,s2)$.
A single structure function, $W_{SA}$, arises from the spin-0 part and 
depends on $F_S$,
and four functions, $W_{SB}$, $W_{SC}$, $W_{SD}$ and $W_{SE}$ are
due to interference between spin-1 and spin-0 amplitudes.

The structure functions can be measured by observing angular distributions
and taking moments $\langle m\rangle $ with respect to products of 
trigonometric functions of these angles.

In the numerical evaluation in Sec.~\ref{num3pi} we will plot
$s_1$, $s_2$ integrated structure functions $w_X$
\begin{eqnarray}
   w_{A,C,SA,SB,SC}(Q^2) & = & \int ds_1\, ds_2\, 
   W_{A,C,SA,SB,SC}(Q^2,s_1,s_2)
\nonumber \\ \nonumber \\
   w_{D,E,SD,SE}(Q^2) & = & \int ds_1 \, ds_2 \, \mbox{sign}(s_1-s_2)
    W_{D,E,SD,SE}(Q^2,s_1,s_2)
\end{eqnarray}
Without the energy ordering $\mbox{sign}(s_1 - s_2)$, the relevant
$w_X$ would vanish due to Bose symmetry.

The integrated decay rate is determined by $W_A$ and $W_{SA}$ only,
the other functions give vanishing contributions after integration 
over the angles. We have
\begin{equation}
   d \Gamma = \frac{G_F^2}{2 M_\tau} \cos^2 \theta_C
   \frac{1}{(4 \pi)^5} \frac{dQ^2}{Q^2} \frac{M_\tau^2 - Q^2}{Q^2}
   \left\{ \frac{1}{2} w_{SA}(Q^2) + \frac{1}{6}
   \left(1 + \frac{2 Q^2}{M_\tau^2} \right) w_{A}(Q^2) \right\}
\end{equation}
%
\section{Calculation of the Hadronic Matrix Elements}

\subsection{Two Pion Decay}
\label{twopion}

The hadronic matrix element which is relevant for the $\tau$ decay into
two pions is the following:
\begin{equation}\label{2pi}
\langle \pi^i(p_1) \pi^l(p_2) {\rm out} | V^k_\mu(0) | 0\rangle =
i \epsilon^{ilk} (p_1-p_2)_\mu F_V(\,s\,) \;\; ,
\end{equation}
where $s= Q^2 = (p_1+p_2)^2$, and $V_\mu^k = \frac{1}{2}
\bar{q}\gamma_\mu\tau^k q$.
In the framework of CHPT, $F_V(\,s\,)$ was calculated by Gasser and
Leutwyler \cite{gasser84} to one loop, 
and by Gasser and Mei{\ss}ner \cite{gassmeiss} up to two loops,
by using a three times subtracted 
dispersive representation. The function to be integrated inside the
dispersive integral is a particular combination of the vector form
factor and the $I=1$, $P$-wave $\pi \pi$ scattering amplitude at tree
and one loop level. In Ref. \cite{gassmeiss} the dispersive integral
was calculated numerically. Here we are able to give a compact
analytic expression of 
this integral \footnote{The integrals one has to calculate here are
similar to the ones that occur in the $\pi \pi$ scattering amplitude
to two loops, see Ref. \cite{pitwoloops_paris,pitwoloops_bern}}:

\begin{equation}\label{fv}
F_V(\,s\,) = 1 + \frac{1}{6}\langle r^2 \rangle_V^{\pi} \,s \,+ c^{\pi}_V 
\, s^2 
+  f_V^{\rm U}\left(\, \frac{s}{M_\pi^2}\,\right)
\end{equation}
\begin{eqnarray}\label{fvu}
 f_V^{\rm U} (\,x\,) &= & 
{\displaystyle \frac{M_{\pi}^2}{16 \pi^2 F_{\pi}^2}}
\left\{{\vrule height1.12em width0em depth1.12em} \right. \!
\frac{x}{9} \left(\,\! 1
 + \,24 \pi^2 \sigma^2
\bar{J}(x) \, \!  \right) - \frac{x^2}{60}
\left.{\vrule height1.12em width0em depth1.12em}  \right\} \nonumber \\
 &+ &
\left({\displaystyle \frac{M_{\pi}^2}{16 \pi^2 F_{\pi}^2}}\right)^2
\left\{ {\vrule height1.12em width0em depth1.12em} \right.
\left[ \bar{l}_2 - \bar{l}_1
+{\displaystyle \frac {\bar{l}_6}{2}}  \,+\,\frac{6 \bar{l}_4}{x} \, \! 
 \right] 
{\displaystyle \frac {x^2}{27}}
\left( \! 1
 + \,24 \pi^2 \sigma^2
\bar{J}(x) \, \!  \right) - {\displaystyle \frac{x^2}{30}}\bar{l}_4
\nonumber
\\
 &+& 
{\displaystyle \frac {3191}{6480}}\,{x}^{2\,} 
+\, {\displaystyle \frac {223}{216}}\,{x}\, -\, 
{\displaystyle \frac {16}{9}}\, 
- \,\frac{ \pi^2\,x}{540} \left(\, 37\, x\, +\, 15\, \right)
 \\
 &+ & {\displaystyle \frac {4 \pi^2}{27}}\,(\,7\,{x}^{2} - 
151\,{x} + 99\,)\,\bar{J}(x)
+ {\displaystyle \frac {2 \pi^2}{9 \, x}}\,
(\,{x}^{3} - 30\,{x}^{2} + 78\,{x} - 128\,) 
\bar{K}_1(x) \nonumber
\\ 
&+ &  8\pi^2 \left( \! {x}^{2}
 - {\displaystyle \frac {13}{3}}\,{x} - 2  \!  \right) \,
\bar{K}_4(x)
\left.{\vrule height1.12em width0em depth1.12em} \right\} \nonumber .
\end{eqnarray}
Where we have used the following functions:
\begin{eqnarray}
\bar{J}(\,x\,) &=& {\displaystyle \frac{1}{16 \pi^2} } 
\left( \, F(\,x\,) + 2 \,\right) \nonumber \; , \\
\bar{K}_1(\,x\,) &=& \frac{1}{16 \pi^2} \frac{F^2(x)}{\sigma^2}
\nonumber \; , \\
\bar{K}_4(\,x\,) &=& \frac{1}{16 \pi^2} \frac{F(x)}
{x \, \sigma^2}
+\frac{1}{32 \pi^2}\frac{1}{x \, \sigma^2} \left[\frac{F^2(x)}
{\sigma^2} 
+ \pi^2 \right]  \\
&+&\frac{1}{48 \pi^2}\frac{1}{x \, \sigma^2}
\left\{\frac{1}{x \, \sigma^4}\left[F^3(x) +
\pi^2 \sigma^2 F(x)\right] - \pi^2 \right\} + 
\frac{1}{192} - \frac{1}{32\pi^2}\ ,
\nonumber
\end{eqnarray}
with
\begin{eqnarray}
F(\,x\,) &=& \sigma \ln {\displaystyle \frac{ \sigma -1}{\sigma+1}}
\; , \\
\sigma &=& \sqrt{1-4/ x} \nonumber \; .
\end{eqnarray}

The functions $\bar{K}_i(\,x\,)$ were recently introduced by
Knecht et al. \cite{pitwoloops_paris}, and like $\bar{J}(\,x\,)$ are
analytic everywhere apart from a branch cut from 4 to $\infty$, and go
to zero as $x \rightarrow 0$.

All the constants which occur in $f_V^{\rm U}(\,s/M_\pi^2\,)$ are known (see
Sec. \ref{CHPT}). The subtraction constants $\langle r^2 \rangle
_V^\pi$ and $c_V^\pi$ are calculable in CHPT and can be expressed in
terms of the low energy constants $l_i^r(\mu)$, chiral logs, and the
new low energy constants which appear in ${\cal L}_6$. With this
representation of the subtraction constants given by CHPT one
automatically satisfies the relevant Ward identities, up to the order
at which one is working. We do not give this 
explicit representation here because up to now there is no information
on the numerical value of the new ${\cal L}_6$ low energy constants.
In the future, with more accurate data on various low energy
processes, and more two loop calculation available, one could try to
pin down at least some of them, but this will require a considerable
amount of work and it is beyond the scope of our analysis.

We adopt in the following the notation of Gasser and Mei{\ss}ner
\cite{gassmeiss} and write 
\begin{eqnarray}\label{rv}
\langle r^2 \rangle _V^\pi &=& {\displaystyle \frac{1}{16 \pi^2
F_\pi^2}} \left[ (\bar{l}_6 -1)+{\displaystyle \frac{ M_\pi^2}{16 \pi^2
F_\pi^2}} \bar{f}_1 \right] + O(M_\pi^4) \nonumber \\
c_V^\pi &=&{\displaystyle \frac{1}{16 \pi^2 F_\pi^2}} \left[
\frac{1}{60 M_\pi^2} + {\displaystyle \frac{1}{16 \pi^2 F_\pi^2}} 
\bar{f}_2 \right] + O(M_\pi^2) \;\; .
\end{eqnarray}
$\bar{f}_1$ and $\bar{f}_2$ contain all the contributions at the two
loop level. In \cite{amendolia} the pion charge radius squared $\langle
r^2 \rangle _V^\pi$ has been determined from experimental data by
means of a simple model for $F_V$ which contains $\langle
r^2 \rangle _V^\pi$ as the only free parameter. The authors obtain the
result $\langle  r^2 \rangle_V^\pi = 0.431 \pm 0.010\, \mbox{fm}^2$,
including the  systematic error as a constraint in the normalization
of the data. We 
repeat the fit with the expression (\ref{fv}) leaving $\langle  r^2
\rangle_V^\pi$ and $c_V^\pi$ as free parameters. Furthermore we include a 
theoretical error, leading to
\begin{eqnarray} \label{eqnfit}
\langle  r^2 \rangle_V^\pi &=& 0.431 \pm 0.020 \pm 0.016 \, \mbox{fm}^2 
\nonumber \\
c_V^\pi &=& 3.2 \pm 0.5\; \pm 0.9 \;\mbox{GeV}^{-4}
\end{eqnarray}
where the first and second errors indicate the statistical and theoretical
uncertainties, respectively.
We reproduce the central value of the radius squared given in
\cite{amendolia} with a larger statistical error, because we fit two
parameters simultaneously: If we keep $c_V^\pi$ fixed, the statistical
error in $\langle 
r^2 \rangle_V^\pi$ reduces by factor of two. The central value of $c_V^\pi$
is rather close to the value obtained by resonance saturation,
$c_V^\pi=4.1\, \mbox{GeV}^{-4}$ \cite{gassmeiss}. 

However we observe that for both parameters the 
theoretical uncertainties are of the same order of magnitude as the
statistical errors. Unless one has a way to keep these theoretical
uncertainties under control, we do not see how $\langle  r^2
\rangle_V^\pi$ can be determined with the accuracy indicated in
\cite{amendolia}. Note that we are not able to fix the low energy
constant $\bar{l}_6$ from our fit, since we do not have independent 
informations on $\bar{f}_1$.   
In the numerical evaluation, we will use the values in (\ref{eqnfit}).

\subsection{Three Pion Decays}
\label{threepi}
As we have seen in Sect. \ref{FF}, it is sufficient to discuss 
only one of the two matrix
elements, the other one can be calculated via the isospin relations
(\ref{isospin}). So we will consider only the one with two neutral
pions.

First of all a general consideration: this matrix element contains a
pole term in $Q^2$ due to the direct coupling of the axial current to
the pion. So the matrix element can be written in general as:
\begin{eqnarray}
\langle \pi^0 (p_1) \pi^0 (p_2) \pi^- (p_3)|
A_\mu^- (0)|0\rangle &=& i \sqrt{2} F_\pi 
\frac{ A_{\pi \pi}(s_3,s_1,s_2) }{M_\pi^2 - Q^2} Q_\mu 
+ \bar{G}(s_1,s_2,s_3) (p_1 + p_2)_\mu  \nonumber \\
          &+& H(s_1,s_2,s_3) (p_1 - p_2)_\mu  
          + \bar{F}(s_1,s_2,s_3) p_{3\,\mu} \; ,
\end{eqnarray}
where $A_{\pi \pi}(s,t,u)$ is the $ \pi \pi$ scattering amplitude as
defined, {\it e.g.} in Ref. \cite{gasser84}.
Note that the separation between the pole term and the barred form
factors $\bar{F}$ and $\bar{G}$ is not unique, however one can split them 
such that the coefficient of the pole is exactly the
$\pi \pi$ scattering amplitude, and therefore define in this way
$\bar{F}$ and $\bar{G}$. 

The calculation of the form factors is done by expanding them in
powers of momenta and quark masses:
\begin{eqnarray}
R &=& i\frac{\sqrt{2}}{F_\pi}\left( R^{(0)} 
      + \frac{R^{(2)}}{F_\pi^2} + \cdots\right)
      \hspace{2cm}R=\bar{F},\bar{G},H \nonumber\\
A_{\pi \pi} &=& A^{(2)}_{\pi \pi} +  A^{(4)}_{\pi \pi} + \cdots 
              \quad ,
\end{eqnarray}
where the superscript $(n)$ indicates a contribution of order $p^n$.
(We remark here that a tree diagram from the Lagrangian ${\cal L}_n$
gives a contribution of order $p^n$ to scattering amplitudes but of
order $p^{n-2}$ to form factors.)

At tree level $H=0$ for the simple reason
that its antisymmetry under exchange of $s_1$ and $s_2$
cannot be satisfied with a constant. As for $\bar{F}^{(0)}$ and
$\bar{G}^{(0)}$, a constant satisfies their symmetry properties 
(\ref{symm}), and we find 
\begin{eqnarray}
\bar{F}^{(0)}&=& - 1 \nonumber \\
\bar{G}^{(0)}&=& 1 \; .
\end{eqnarray}
At the one loop level we have the following results:
\begin{eqnarray}
\bar{F}^{(2)}&=&\frac{1}{3}M_\pi^2 [\bar{J}(\hat{s}_1) 
                 + \bar{J}(\hat{s}_2)]
                 - \frac{1}{12}(s_1 - s_2) [\bar{J}(\hat{s}_1)
                 - \bar{J}(\hat{s}_2)]
                 - \frac{1}{2} s_3 \bar{J}(\hat{s}_3) \nonumber \\ 
&& + \frac{1}{96\pi^2}\left[-2\bar{l}_1(s_3 - 2M_\pi^2) 
                         + \bar{l}_2(s_1 + s_2 + s_3 - 4M_\pi^2)
                         - 6\bar{l}_4M_\pi^2\right. \nonumber \\ 
    &&\left.\hspace{1.5cm} - \bar{l}_6(s_1 + s_2 + 2s_3 - 4M_\pi^2)
                         - \frac{1}{2}(s_1 + s_2 - 5s_3) 
                         + \frac{8}{3}M_\pi^2\right] \nonumber \\[2mm]
\bar{G}^{(2)}&=&-\frac{1}{6}M_\pi^2 [\bar{J}(\hat{s}_1) 
                 + \bar{J}(\hat{s}_2)]
                 - \frac{1}{12}(s_1 - s_2) [\bar{J}(\hat{s}_1) 
                 - \bar{J}(\hat{s}_2)]
                 + \frac{1}{2} s_3 \bar{J}(\hat{s}_3) \nonumber \\ 
&& + \frac{1}{96\pi^2}\left[2\bar{l}_1(s_3 - 2M_\pi^2) 
                         - \bar{l}_2(s_1 + s_2 - s_3 - 4M_\pi^2)
                         + 6\bar{l}_4M_\pi^2\right. \nonumber \\
    &&\left.\hspace{1.5cm} + \bar{l}_6(s_1 + s_2 - 2M_\pi^2)
                         + \frac{1}{2}(s_1 + s_2 - 7s_3) 
                         - \frac{10}{3}M_\pi^2\right] \nonumber \\[2mm]
H^{(2)}&=&-\frac{1}{6}(s_1 - s_2) 
                                   [\bar{J}(\hat{s}_1) 
                                    + \bar{J}(\hat{s}_2)]
                       - \frac{1}{6}(s_1 + s_2 - 5M_\pi^2) 
                                   [\bar{J}(\hat{s}_1) 
                                    - \bar{J}(\hat{s}_2)] \nonumber \\
&& + \frac{1}{96\pi^2}\left[-2\bar{l}_2(s_1 - s_2) 
                         + \frac{5}{3}(s_1 - s_2)\right] \; \; ,
\end{eqnarray}
where $\hat{s}_i=s_i/M_\pi^2$, and the $\bar{l}_i$ are listed in Tab.~1.

As for $A_{\pi \pi}$ its expansion is now known up to the two loop level
\cite{pitwoloops_bern}. Here we need only the first two terms of the
expansion which are known since a long time \cite{weinpipi,gasser84}:
\begin{eqnarray}
F_\pi^2 A_{\pi \pi}^{(2)}(s,t,u) &=& s-M_\pi^2 \nonumber \\
F_\pi^4 A_{\pi \pi}^{(4)}(s,t,u) &=&        
           \frac{1}{12}[(t - u)^2 - 2M_\pi^2 s + 4M_\pi^4]
                         [\bar{J}(\hat{t}) + \bar{J}(\hat{u})] \nonumber \\  
         && - \frac{1}{12}(t - u) (s + 2M_\pi^2)
                         [\bar{J}(\hat{t}) - \bar{J}(\hat{u})]
           + \frac{1}{2} (s^2 - M_\pi^4) \bar{J}(\hat{s}) \nonumber \\ 
   &&+ \frac{1}{96\pi^2}\left\{2\bar{l}_1(s - 2M_\pi^2)^2 
                          + \bar{l}_2[(t - u)^2 + s^2]\right. \nonumber \\
   &&\left.\hspace{1.5cm} - 3\bar{l}_3 M_\pi^4
                          + 12\bar{l}_4 M_\pi^2(s - M_\pi^2)
                          - \frac{5}{6}(t - u)^2 \right. \nonumber \\
   &&\left.\hspace{1.5cm} - \frac{7}{2}s^2 
                          - \frac{1}{3}M_\pi^2(4s - 13M_\pi^2)\right\} \; .
\end{eqnarray}                         
The form factors $F$ and $G$ as defined in (\ref{decomp}) can be easily
reconstructed from $A_{\pi  \pi}$ and the corresponding barred functions 
at each given order, via the simple relation:
\begin{eqnarray}
R &=& i\frac{\sqrt{2}}{F_\pi}\left( R^{(0)} 
      + \frac{R^{(2)}}{F_\pi^2} + \cdots\right)\nonumber\\
R^{(n-2)} &=& \frac{F_\pi^n\,A_{\pi \pi}^{(n)}}{M_\pi^2 - Q^2} +
\bar{R}^{(n-2)} \hspace{2cm}R=F,G \; .
\end{eqnarray}
One could wonder whether it is possible to experimentally disentangle
the contribution of $A_{\pi \pi}$ to some of these form factors. We
are convinced that this is not the case. The reason being that $A_{\pi \pi}$
contributes only to $F_S$, which is very difficult to measure,
 and that there are in addition other contributions to $F_S$. 
%
\section{Numerical Results for the Branching Ratios and the 
Structure Functions}
\label{decayrate}
\subsection{Two Pion Decay}
\begin{table}  
   \caption{Integrated branching ratios for 
\protect{$\protect \sqrt{Q^2} 
\leq Q_{max}
$} \label{tabrates}
 predicted
 with CHPT at a given order 
\protect{$O(p^n)$} 
and from vector meson dominance models (VMD), see text.}
$$
   \begin{array}{cccccc}
   \hline \mbox{ } \\
   \mbox{mode} & Q_{max} [\mbox{MeV}] &
   O(p^2) & O(p^4) & O(p^6) & \mbox{VMD} \\ \mbox{ } \\ 
   \hline \mbox{ } \\ 
   & 400 & 4.22 \times 10^{-4} & 6.66 \times 10^{-4} & 7.34 \times 10^{-4} & 
7.40 \times 10^{-4} \\
   \tau \to 2 \pi \nu_\tau
   & 500 & 1.57 \times 10^{-3} & 2.92 \times 10^{-3} & 3.45 \times 10^{-3} & 
3.72 \times 10^{-3} \\
   & 600 & 3.40 \times 10^{-3} & 7.57 \times 10^{-3} & 9.73 \times 10^{-3} & 
1.20 \times 10^{-2} \\
   \mbox{ } \\ 
   \hline \mbox{ } \\ 
   & 500 & 2.00 \times 10^{-7} & 4.19 \times 10^{-7} & & 4.19 \times
10^{-7} \\ 
   \tau \to 3 \pi \nu_\tau
   & 600 & 4.31 \times 10^{-6} & 1.07 \times 10^{-5} & & 1.23 \times
10^{-5} \\ 
   & 700 & 2.21 \times 10^{-5} & 6.55 \times 10^{-5} & & 9.51 \times
10^{-5} \\ 
   \mbox{ } \\ 
   \hline \hline
   \end{array}
$$
\end{table}
At first, let us discuss integrated branching ratios 
$\mbox{BR}_{2\pi}(Q_{max})$
\begin{equation}
 \mbox{BR}_{2\pi}(Q_{max}) = \mbox{BR}_{e}
   \int_{4 M_\pi^2}^{Q_{max}^2} dQ^2 \frac{d \Gamma_{2 \pi}}{\Gamma_e 
   \, d Q^2}(Q^2)
\end{equation}
The results from CHPT are given in Tab.~\ref{tabrates}.
{}From the convergence of the expansion in $p^2$ we conclude that the
CHPT expansion truncated at this order works fine up to $Q_{max} =
500\,\mbox{MeV}$.  

We also compare with
the prediction from a VMD model, viz. from the model 1 of \cite{Kue90},
which has been implemented in TAUOLA \cite{tauola}.
This VMD model parameterizes $F_V$ in terms of a coherent superposition 
of a $\rho$ and a $\rho'$ Breit-Wigner, with an overall normalization 
fixed by matching to the $O(p^2)$ chiral prediction.
It gives a good parameterization of $e^+ e^- \to 2 \pi$ annihilations
in the range covered by the tau mass.
We find that in the range up to $500$ or $600\,\mbox{MeV}$, where
we trust CHPT, the predictions of CHPT and from the VMD model agree well.

\begin{figure}                                                 
\caption{Differential decay rate for $\tau \to 2 \pi \nu_\tau$: Predictions
by CHPT at $O(p^2)$ (dashed), at $O(p^4)$ (dashed-dotted), at $O(p^6)$ 
(solid),
and from a vector meson dominance model (dotted), compared with
experimental data from CLEO (dots with error bars)}
\label{fig2pi}
\bildchen{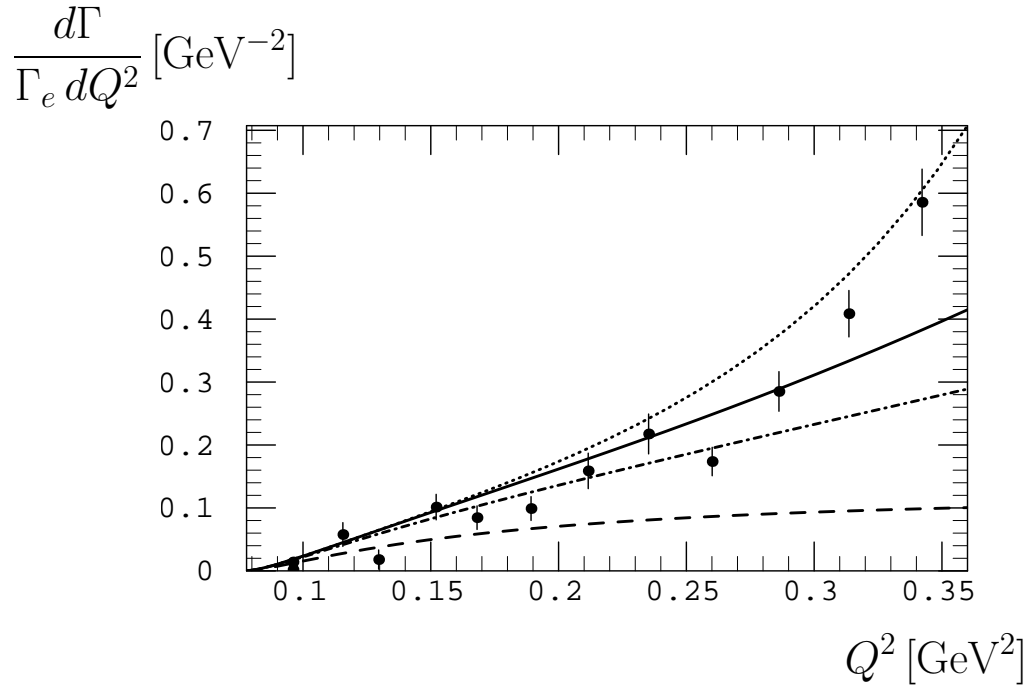}{\frac{d \Gamma}{\Gamma_e \,dQ^2}\,[\mbox{GeV}^{-2}]}%
{Q^2\,[\mbox{GeV}^2]}
\end{figure}
In Fig.~\ref{fig2pi} we plot the two pion invariant mass spectrum
normalized to the electronic branching ratio of the tau. We plot 
the predictions from CHPT, together with the prediction from the VMD
model in \cite{Kue90} and preliminary data from CLEO \cite{cleo}.
Note that we used a simplified approach to fix the overall
normalization of the data in \cite{cleo}. We multiplied the spectrum from
\cite{cleo} with a
normalization factor $N$ and determined $N$ by fitting 
the data to the VMD prediction of \cite{Kue90}.
Of course, the normalization should instead be taken from the data.
In fact we suggest a careful reanalysis of the low energy part
of the spectrum and its absolute normalization in order to compare it with
the CHPT prediction.

It is of some interest to understand how sensitive 
$\tau\to 2\pi\nu_\tau$ decays are to the pion charge
radius 
$\langle r^2\rangle _V^\pi$, which is defined from the expansion of
$F_V(Q^2)$  in terms of $Q^2$ 
\begin{equation}
  F_V(Q^2) = 1 + \frac{1}{6} \langle r^2\rangle _V^{\pi} Q^2 + c_V^\pi 
  Q^4 + f_V^{\rm U}\left(Q^2\right) + O(Q^6) 
\end{equation}
where $f_V^{\rm U}(Q^2)$ is given in (\ref{fvu}) and is very small
numerically. 
According to the previous results, we 
consider this expansion valid up to $Q_{max} =
500\,\mbox{MeV}$. Furthermore we neglect theoretical uncertainties due to
higher order corrections and assume that $c_V^\pi$ is known exactly. For a
discussion of these points see Sec.\ref{twopion}\,.  

Given a number $N$ of events $\tau\to
2 \pi \nu_\tau$ with hadronic invariant mass squared $Q^2$ in the interval
$4 M_\pi^2 \cdots Q^2_{max}$,  the precision with which 
$\langle r^2 \rangle_V^\pi$ can be measured is (see \cite{franzi})
\begin{equation}                                      \label{eqnfranzi}
   \sigma_p = \frac{1}{\sqrt{N}} \left[
   \int \frac{1}{f} \left( \frac{\partial f(x;p)}{\partial p}
   \right)^2 \, dx \right]^{-1/2} 
 = 
 \frac{1}{\sqrt{N}} 
   \, 7.37 \, \mbox{fm}^2
\end{equation}
where
\begin{equation}
  f(Q^2,\langle r^2\rangle _V^\pi) := \frac{1}{R} \, \rho(Q^2)
\end{equation}
with
\begin{eqnarray}
  \rho(Q^2) = \frac{1}{\Gamma_e} \frac{d\Gamma_{2\pi}(Q^2)}{dQ^2}
\nonumber \\ \nonumber \\
  R = \int_{4 M_\pi^2}^{Q^2_{max}} dQ^2 \rho(Q^2)
\end{eqnarray}

So given $N_\tau$ decaying taus and an detection efficiency $\eta$, 
we have 
\begin{equation}
   N = N_\tau\,  \eta \, 
   \mbox{BR}(\tau \to 2\pi \nu_\tau,\,Q^2 \leq 0.25 \,\mbox{GeV}^2)
\end{equation}
where the branching ratio 
$\mbox{BR}(\tau \to 2\pi \nu_\tau,\,Q^2 \leq 0.25 \,\mbox{GeV}^2) 
= 3.6 \times 10^{-3}$ according to Tab.~\ref{tabrates}

Based on this, we now give rough order of magnitude estimates for 
the possible statistical accuracy of present and future experiments.
We assume an efficiency of $\eta = 30 \%$.
CESR has at present about
$5 \times 10^6$ taus, thus the possible statistical accuracy is
of the order of $\sigma_{\langle r^2\rangle _V^\pi} = 0.1 \, \mbox{fm}^2$. 
A b-factory might have $5 \times 10^7$ taus per year.  
Assuming 3 years of running time, this leads with the assumptions mentioned
above to a possible  statistical accuracy of the order of
$\sigma_{\langle r^2\rangle _V^\pi} = 0.02 \, \mbox{fm}^2$.

These numbers have to be compared to the accuracy evaluated from $ \pi e$
scattering. With the assumptions we made above, the
present result is $\langle r^2 \rangle_V^\pi
= 0.431 \pm 0.010 \, \mbox{fm}^2$ \cite{amendolia}.  
So it seems difficult for tau decays to become competitive
with $\pi e$ scattering for the determination of $\langle r^2
\rangle_V^\pi$.  
Nevertheless, tau decays can provide an interesting cross-check.
%
\subsection{Three Pion Decay}
\label{num3pi}
\begin{figure}                                                  
\caption{Integrated structure function $w_A(Q^2)$ for $2\pi^- \pi^+$:
CHPT prediction at $O(p^2)$ (dashed), $O(p^4)$ 
(solid) and from a vector meson dominance model (dotted).
The four functions $w_A$ and $w_C$ for both modes $2\pi^- \pi^+$
and $2 \pi^0 \pi^-$ all look identical within the resolution of this
diagram.}
\label{figwa}
\bildchen{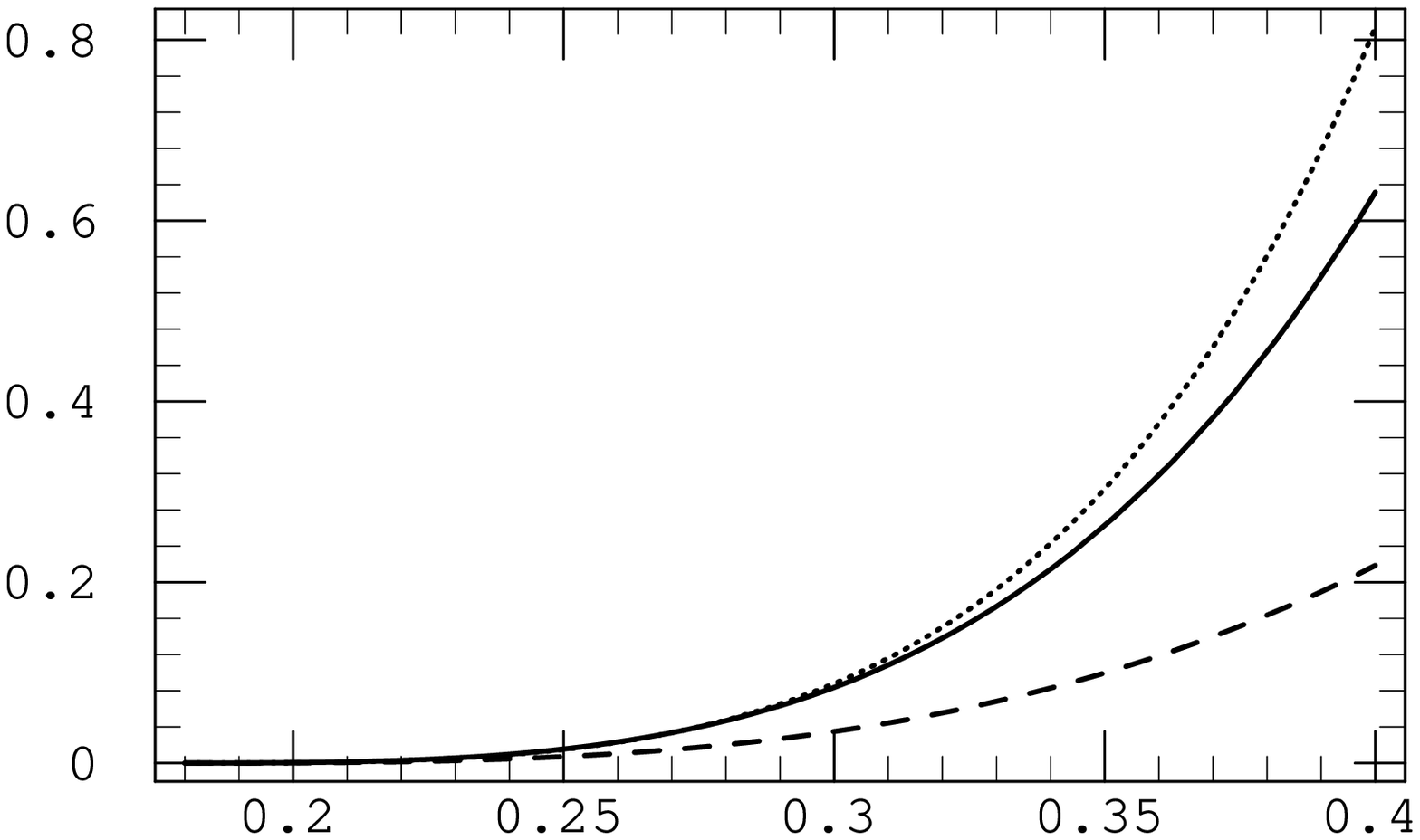}{w_A [\mbox{GeV}^{4}]}%
{{Q^2}\,[\mbox{GeV}^2]}
\end{figure}
\begin{figure}           
\caption{Integrated structure function $w_D(Q^2)$. 
CHPT at $O(p^4)$ for $2\pi^-\pi^+$ (solid) and for $2\pi^0\pi^-$
(dashed-dotted) and form the VMD model (dotted, identical prediction
for both charge modes)}  
\label{figwd}
\bildchen{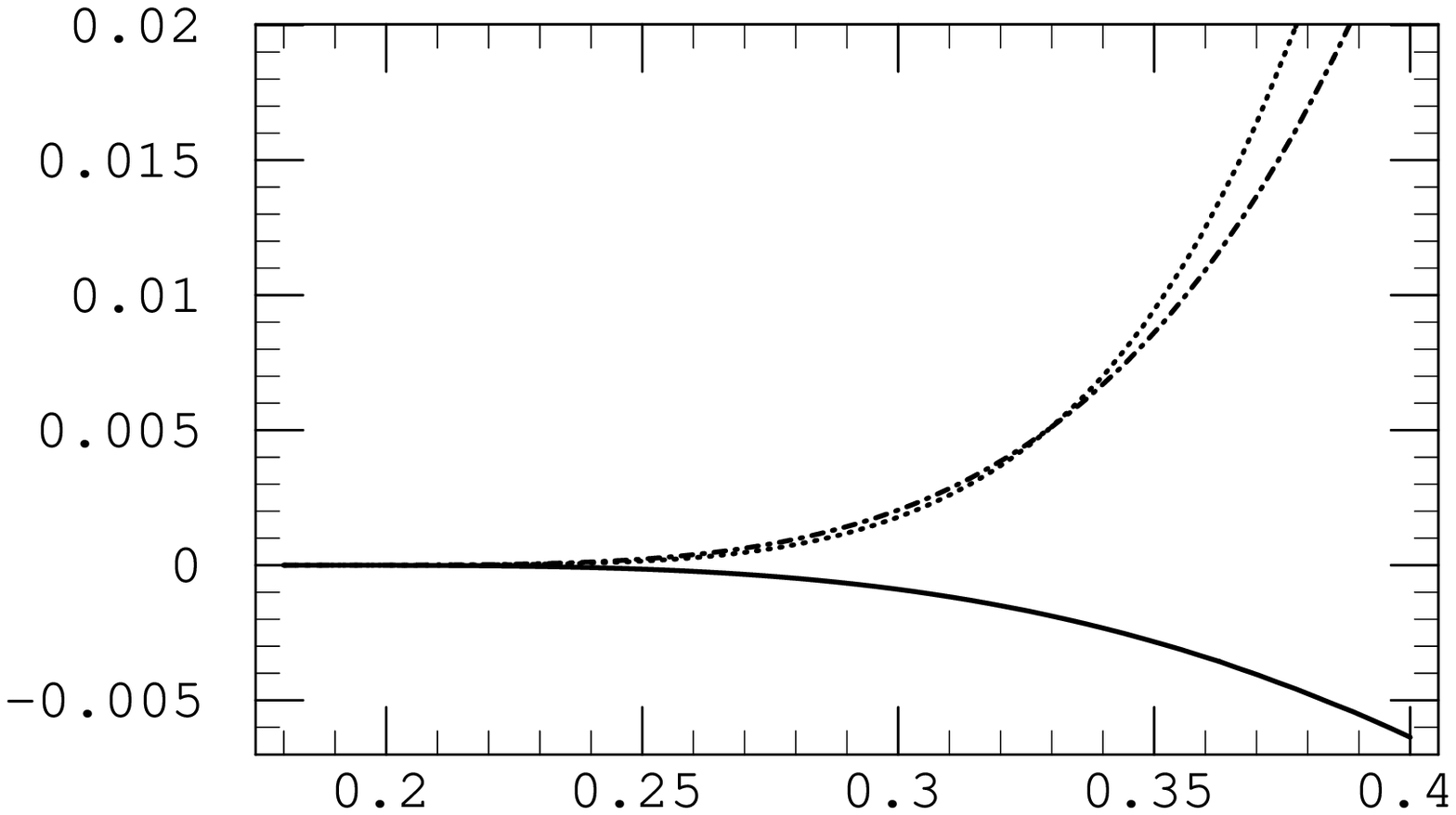}{w_D [\mbox{GeV}^{4}]}%
{{Q^2}\,[\mbox{GeV}^2]}
\end{figure}
\begin{figure}                                                  
\caption{Integrated structure function $w_E(Q^2)$. 
CHPT at $O(p^4)$ for $2\pi^-\pi^+$ (solid) and for $2\pi^0\pi^-$
(dashed-dotted) and from the VMD model (dotted, identical prediction
for both charge modes) }
\label{figwe} 
\bildchen{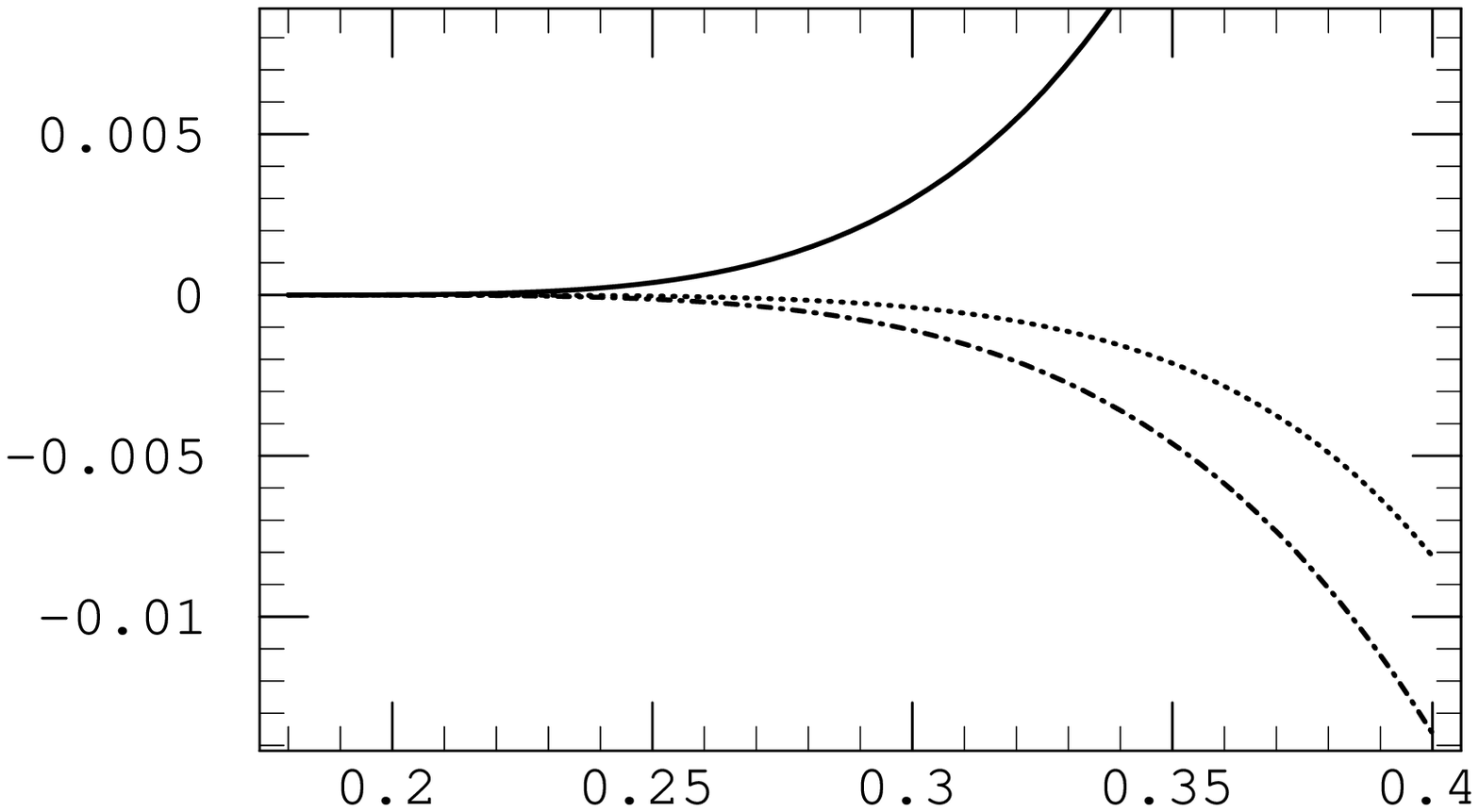}{w_E [\mbox{GeV}^{4}]}%
{{Q^2}\,[\mbox{GeV}^2]}
\end{figure}
\begin{figure}                                                  
\caption{Integrated structure function $w_{SA}(Q^2)$
for $2\pi^-\pi^+$. CHPT prediction at $O(p^2)$ (dashed) and
$O(p^4)$ (solid), and from a VMD model (dotted).}\label{figwsaP}
\bildchen{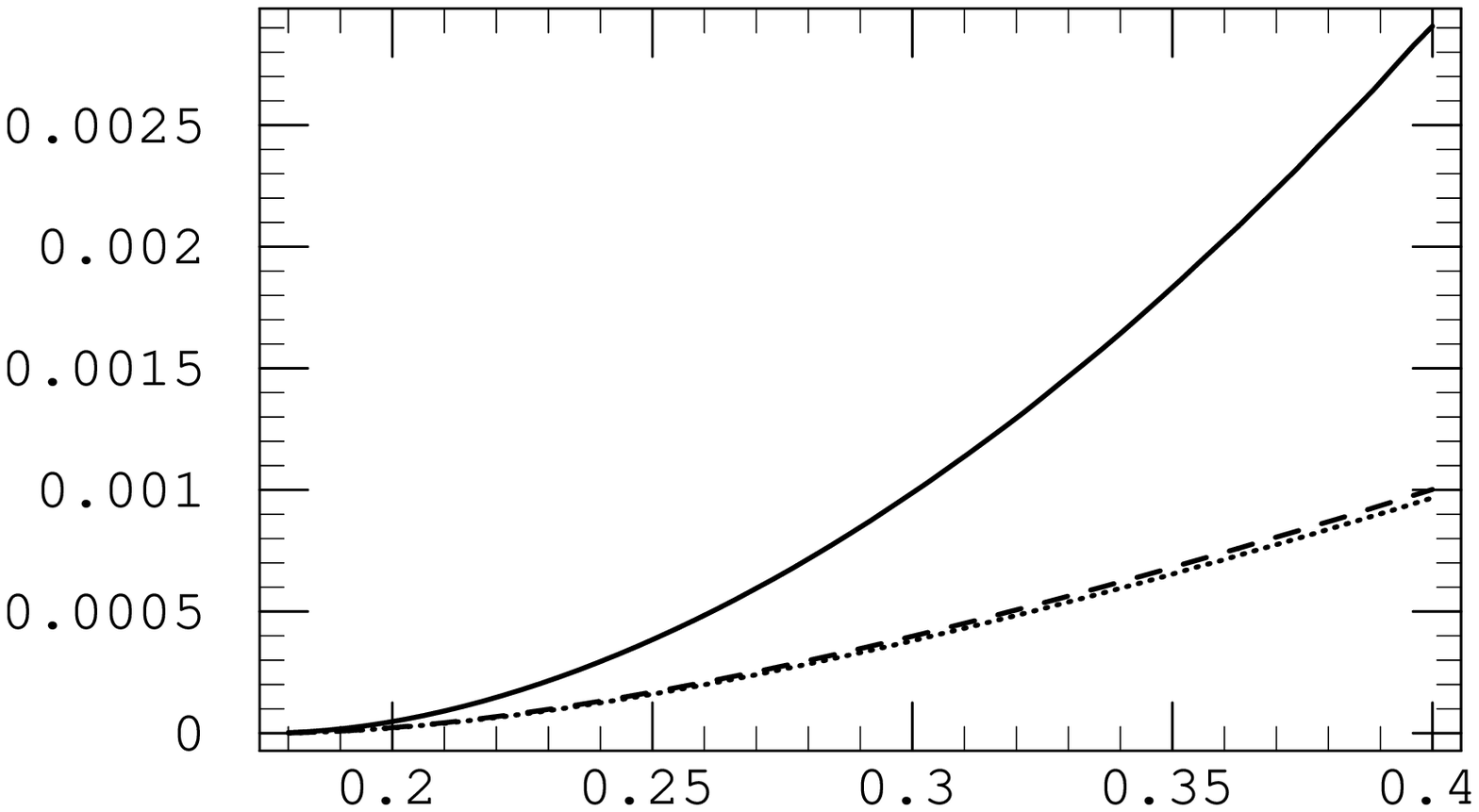}{w_{SA} [\mbox{GeV}^{4}]}%
{{Q^2}\,[\mbox{GeV}^2]}
\end{figure}
\begin{figure}                                                  
\caption{Integrated structure function $w_{SA}(Q^2)$
for $2\pi^0\pi^-$. CHPT prediction at $O(p^2)$ (dashed) and
$O(p^4)$ (solid), and from a VMD model (dotted).}\label{figwsa0}
\bildchen{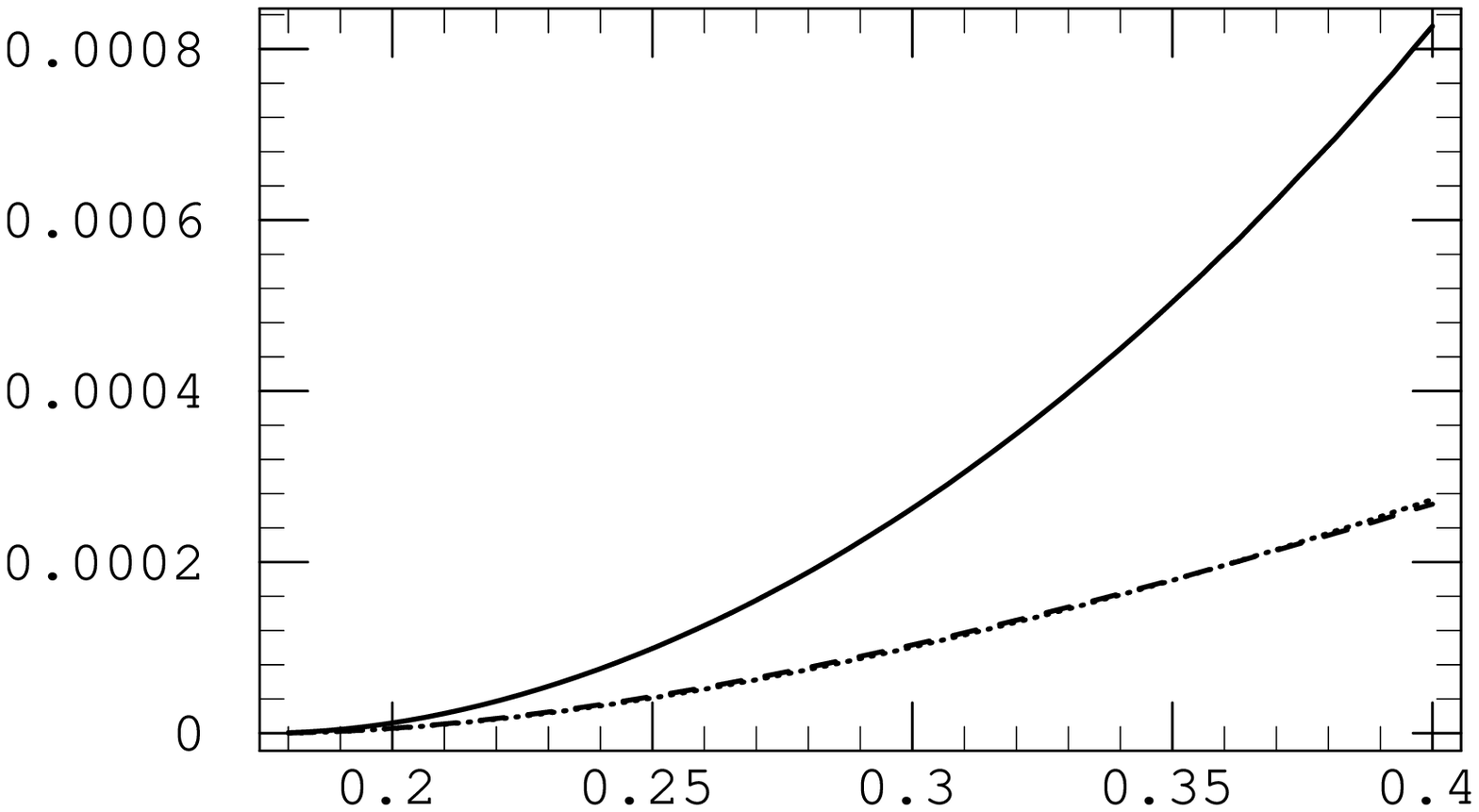}{w_{SA} [\mbox{GeV}^{4}]}%
{{Q^2}\,[\mbox{GeV}^2]}
\end{figure}
\begin{figure}                                                  
\caption{Integrated structure functions $w_{SB}(Q^2)$.
Prediction from CHPT for $2\pi^- \pi^+$ (solid), $2\pi^0\pi^-$
(dashed-dotted), 
and from a VMD model for  $2\pi^- \pi^+$ (dotted), $2\pi^0\pi^-$
(dashed)} \label{figwsb}
\bildchen{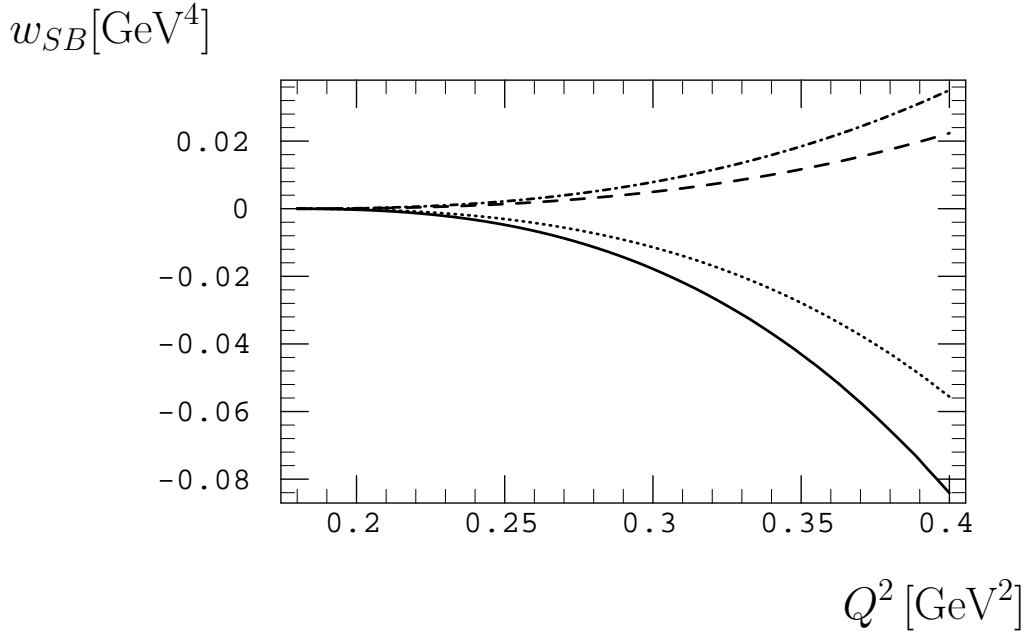}{w_{SB} [\mbox{GeV}^{4}]}%
{{Q^2}\,[\mbox{GeV}^2]}
\end{figure}
\begin{figure}
\caption{Dalitz plot distribution of the $2\pi^- \pi^+$ final
state in $s_1$, $s_2$ for $Q^2=0.36\,\mbox{GeV}^2$.}
\label{figdiff}
\bildchenp{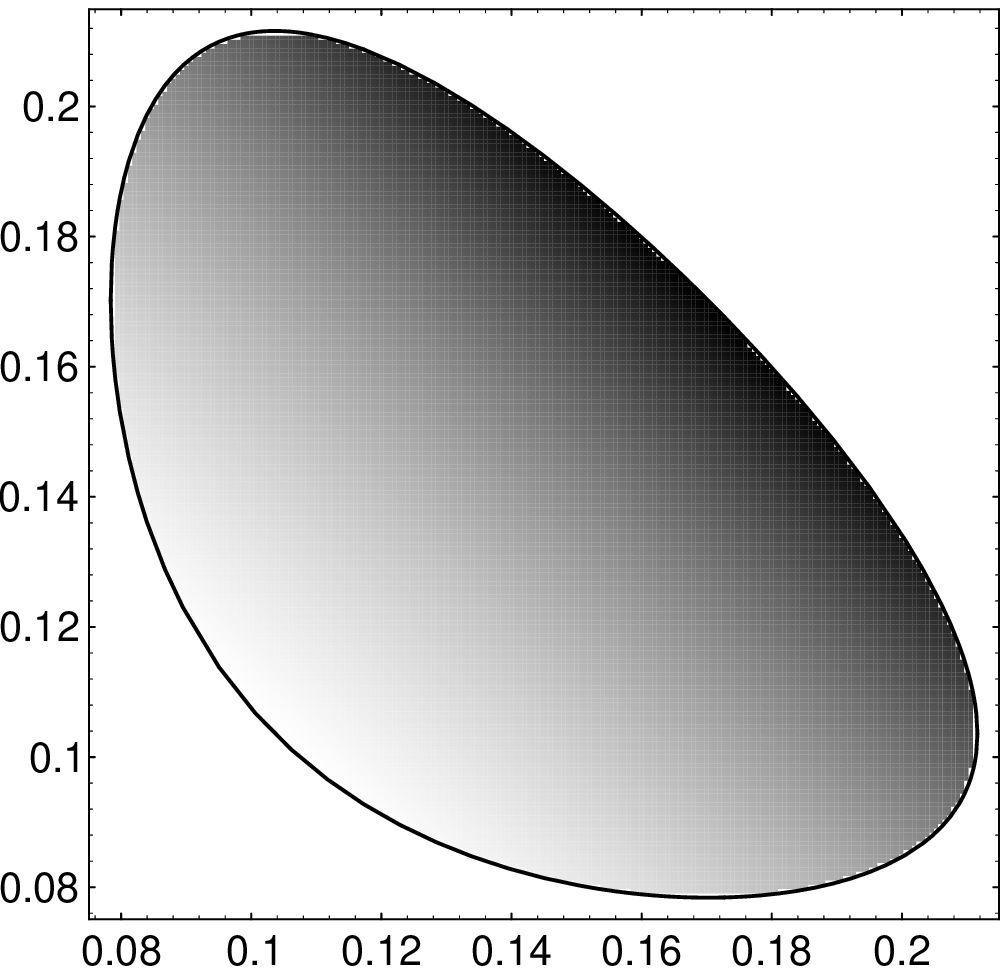}{s_1\,[\mbox{GeV}^2]}{s_2\,[\mbox{GeV}^2]}
\end{figure}
The numerical results for the integrated branching ratio with $Q^2
\leq Q^2_{max}$ for CHPT at $O(p^2)$ and $O(p^4)$ are given in
Tab.~\ref{tabrates}. The series expansion of CHPT does not seem to
behave very well. Even for $Q_{max}=500\,\mbox{MeV}$,  the
integrated one loop contribution is already around 45 \% of the tree
level result. This means that to have a CHPT prediction with a
reasonably small theoretical uncertainty for such a quantity one
should stop at $Q_{max}$ well below $500\, \mbox{MeV}$, and this
would be very difficult to test experimentally, because of the strong
phase space suppression of this region. On the other hand we observe
that there is a fair agreement with the VMD model numbers up to
$Q_{max}=600 \,\mbox{MeV}$, which is better than what happens in
the two pion case.
This can be understood in terms of the fact that in the two pion case,
the $\rho(770)$ resonance is very close, whereas in the three pion
case, the nearest three pion resonance, the $a_1(1260)$, is much
further away.
Whether the VMD model is a good representation of the experimental
data even at such a low energy though, has still to be verified.

Beyond the integrated decay rates, this decay mode has a very
rich structure, which in principle could be investigated in detail
experimentally. In fact in \cite{KM2} it has been shown
that it is possible to extract all three form factors from a
measurement of all angular distributions. One could then
compare the measured form factors to the analytic formulae of CHPT. 
This extraction is however very difficult in practice, and moreover the
phase space suppresses considerably the region where the comparison
makes sense. For these reasons we find more useful to show directly
the curves for the integrated structure functions $w_X$ in
Figs.~\ref{figwa}--\ref{figwsb}. 
We have chosen six out of the nine structure functions 
which are present, mainly because the missing three are too difficult to
be measured and not particularly interesting.
Note that, in accordance with our definitions in 
Eqs.~(\ref{cca},\ref{ccb}), the Cabibbo angle 
$\cos^2 \theta_C$ is factored out from hadronic matrix element, and
so the structure functions $w_X$ do not include this factor.

In Fig.~\ref{figwa} we plot $w_A(Q^2)$ for the $2 \pi^- \pi^+$ final state.
It turns out that $w_A(Q^2)$ and $w_C(Q^2)$ for both charge states,
$2 \pi^- \pi^+$ and $2 \pi^0 \pi^-$ are all very similar, so the
four plots can not be distinguished from each other. 
The equality of the structure functions for the two charge modes suggests
that $w_A$ and $w_C$ are dominated by the [210] partition.
We have explicitly verified that this is the case: for example at $Q^2
= 15 M_\pi^2$ the $[210]$ partition contributes $6.73 \times 10^{-2}
\mbox{GeV}^{-4}$ to $w_A$ for both modes, whereas the $[300]$
partition contributes $2.21 \times 10^{-4} \mbox{GeV}^{-4}$ for the $2
\pi^- \pi^+$ and, according to Eq.~(\ref{eqn30}), one quarter of this for
the $2 \pi^0 \pi^-$ mode. The 
interference of the two partitions is completely negligible.

The reason why $w_A \sim w_C$ at low energy is that
\begin{eqnarray}
W_C &=& W_A - 2 x_3^2 | F_1-F_2 |^2 \nonumber \\
    &=& W_A - 8 x_3^2 | H |^2 \; .
\end{eqnarray}
Since $H$ is zero at tree level, the difference vanishes at leading
order. Moreover this difference starts as the square of a quantity of
$O(p^2)$. Using the language of the sixties we can say that the vanishing
of this difference near threshold is a Low Energy Theorem (LET), and that
it receives corrections only at next--to--next--to--leading order.
We also notice that $H$ is a function antisymmetric under exchange of
$s_1$ and $s_2$, so that its modulus squared has a zero along the line
$s_1=s_2$. At low energy, where the distance between this line and the
boundaries of integration is not large (in units of $M_\pi^2$), the
presence of the zero produces an additional suppression of the integral.
The similarity of these two functions near threshold
was already found in Ref. \cite{KM1,KM2} in the framework of a VMD model.
Here however, we can give a detailed algebraic account of why this
happens. For all these arguments we believe that this fact should be
verified experimentally even at energies well above those
where one would trust a one loop CHPT calculation.

In Fig.~\ref{figwd} and \ref{figwe} we plot $w_D(Q^2)$ and $w_E(Q^2)$.
These are much smaller than $w_A$ near threshold, and certainly more
difficult to be measured. However they seem rather interesting from a
theoretical point of view. First of all for both of them the $O(p^2)$
contribution vanishes.
Secondly, the $O(p^4)$ predictions from CHPT differ strongly for the
two charge modes. 
According to the discussion in Sec.~\ref{secpartitions}, this fact
suggests that near threshold these two structure functions are
strongly influenced by the partition $[300]$. 
Using the decomposition of the form factors given in Eq.~(\ref{ffpart}),
we have verified that this is exactly what happens. As can be seen in
Tab.~\ref{wed} the change of sign in the interference contribution is
responsible for the change of sign of the whole integrated structure
functions in the two different charge modes.
\begin{table}
\begin{center}
 \begin{tabular}{|lr|c|c|c|c|}
   \hline  
\mbox{}&\mbox{}& \mbox{} & \mbox{} & \mbox{} & \mbox{} \\
 \mbox{}& \mbox{}& $[300]$ & $[210]$ & interf. & total \\
\hline
\mbox{}&\mbox{}& \mbox{} & \mbox{} & \mbox{} & \mbox{} \\
\mbox{}&$ (2\pi^- \pi^+)$& $8.63 \times 10^{-2}$ & $1.02 $ &$-1.56$&
$-0.450$ \\ 
$10^3 w_D $& \mbox{}&\mbox{} & \mbox{} & \mbox{} & \mbox{} \\
\mbox{}&  $(2\pi^0 \pi^-)$&$2.16 \times 10^{-2}$ & $1.02$ &$0.779$&$1.82$ \\
\mbox{}&\mbox{}& \mbox{} & \mbox{} & \mbox{} & \mbox{} \\
\hline
\mbox{}&\mbox{}& \mbox{} & \mbox{} & \mbox{} & \mbox{} \\
\mbox{}&$ (2\pi^- \pi^+)$& $4.62\times 10^{-4}$ & $0.206 $ &$2.08$ &$2.29$ \\
$10^3 w_E$& \mbox{}&\mbox{} & \mbox{} & \mbox{} & \mbox{} \\
\mbox{}& $(2\pi^0 \pi^-)$&$1.16\times 10^{-4}$ & $0.206$&$-1.04$&$-0.837$ \\
\mbox{}&\mbox{}& \mbox{} & \mbox{} & \mbox{} & \mbox{} \\
\hline
 \end{tabular}
\end{center}
\caption[]{Splitting of the integrated structure functions $w_E$ and
$w_D$ into the contributions of the partitions $[300], \; [210]$ and
their interference, at $Q^2 = 15 M_\pi^2$. }
\label{wed}
\end{table}
As far as we know up to now there are no data for these structure
functions so close to threshold: so this is a real prediction of CHPT.
Moreover this change of sign is absent in all the models of which we
are aware which have been used to describe this decay channel.
It is then important to ask how reliable this prediction is and up to
what energy it can be trusted. These questions are especially
difficult to answer here because we have only a leading order
calculation (since the $O(p^2)$ contribution vanishes).
The only thing we can do is to check how sensitive the prediction is
to the values we use for the low energy constants. We concentrate here
on the $[300]$ state, since it is the one responsible for the new effect:
\begin{eqnarray}
\frac{F_\pi^3}{i \sqrt{2}} F_1^{[300]} &=& \frac{1}{72 \pi^2 } \left[
\bar{l}_1+\bar{l}_2 -3 
\right] \left(\frac{Q^2}{3}-s_1+M_\pi^2 \right) \nonumber \\
&+&\frac{4}{9}\bar{J}(s_1)(M_\pi^2-s_1) -
\frac{2}{9}\bar{J}(s_2)(s_2-M_\pi^2) -
\frac{2}{9}\bar{J}(s_2)(s_2-M_\pi^2) \; .
\end{eqnarray}
First of all we notice that this partition has no contribution from
tree level. This is not the case for the $[210]$ partition, which
starts as:
\begin{equation} 
\frac{F_\pi}{i \sqrt{2}} F_1^{[210]}= \frac{2}{3}+O(p^2) \; .
\end{equation}
This explains why the interference contribution is much bigger than
the one from the $[300]$ partition alone.
Then we may easily see from the definition of these two structure
functions given in App.~\ref{app2} that the $w_D$ is mainly sensitive to the
real part of $F_1^{[300]}$, whilst $w_E$ to the imaginary part. This
means that only the numbers for $w_D$ may depend on the combination of
low energy constants which occurs in $F_1^{[300]}$.
The value used for the combination $\bar{l}_1+\bar{l}_2-3$ in the
numerical calculations is $1.4$, resulting from the central values
given in Tab.~\ref{li}. The uncertainty on this number can be
estimated from the same Tab.~\ref{li} to be $\pm 1.1$. So the
possibility that the real value of this combination be twice as much,
or very close to zero cannot be excluded. We have checked that by
changing the value of $\bar{l}_1+\bar{l}_2-3$ by $\pm 1.4$ changes
$w_D$ in the $2 \pi^- \pi^+$ mode by $\pm 4.4 \times 10^{-4}$, at
$Q^2=15 M_\pi^2$.
As expected $w_E$ remains practically unchanged.
Even in the worst case for $w_D$, however, there remains a sizable
difference between the two charge modes.
The effect of higher orders in the chiral expansion and the related
questions of how far in energy one can trust this prediction will
remain unanswered until one calculates the form factors at two loops,
which is beyond the scope of our analysis. We may just argue that
since there are no low-lying 
resonances contributing to this particular three pion
state, one of the possible sources of large higher order corrections
is excluded.

All in all we can say that CHPT predicts a sign difference in the
two structure functions $w_D$ and $w_E$ for the two charge modes near
threshold. 
This prediction will have interesting consequences in the next subsection,
where we compare with VMD models and available data.

In Fig.~\ref{figwsaP} we plot $w_{SA}(Q^2)$ for the $2\pi^- \pi^+$ 
state. The corresponding curves for the $2\pi^0 \pi^-$ state in
Fig.~\ref{figwsa0} have a very
similar shape but a different overall normalization. In fact,
the ratio $w_{SA}^{2\pi^-\pi^+} / w_{SA}^{2\pi^0\pi^-}$ is close to 
4, which according to Sec.~\ref{secpartitions} implies 
that the scalar form factor at low energies is dominated by the
[300] state. 

We find that the structure function $w_{SA}$ is very small  compared
to $w_{A}$, so the spin-0 contribution to the integrated rate is
negligible. The only chance to measure the spin-0 part is via its
interference with the spin-1 part in the structure function 
$w_{SB}(Q^2)$, which is plotted in
Fig.~\ref{figwsb}.

In Fig. \ref{figdiff} we consider the Dalitz plot distribution $d \Gamma/(d
Q^2\, ds_1\,ds_2)$ in $s_1$, $s_2$ for fixed $Q^2=0.36\,\mbox{GeV}^2$. We
display the $O(p^4)$ prediction for the $2\pi^-\pi^+$ mode.  
At $O(p^2)$, the Dalitz plot density depends only on $s_1 + s_2$.
As seen from the figure, this feature seems to persist at $O(p^4)$.
Let us analyze this issue quantitatively.
To describe the Dalitz plot, we define $s_+$ and $s_-$ by
\begin{equation}
   s_+ := s_1 + s_2  , \qquad
   s_- := s_1 - s_2
\end{equation}
and then replace $s_+, s_-$ by dimensionless variables $x,y$ 
with $0 \leq x \leq 1$ and $-1 \leq y \leq 1$ via
\begin{eqnarray}
   s_+ & = & (s_+^{\rm max} - s_+^{\rm min}) x + s_+^{\rm min}
\nonumber \\
\nonumber \\
   s_- & = & s_-^{\rm max}  y
\end{eqnarray}
where
\begin{eqnarray}
   s_+^{\rm min} & = & 2 m_\pi (m_\pi + \sqrt{{Q^2}})
\nonumber \\
\nonumber \\
   s_+^{\rm max} & = & {Q^2} - m_\pi^2
\nonumber \\
\nonumber \\
   s_-^{\rm max} & = &  \sqrt{\frac{[s_+ (s_+ - 4 m_\pi^2) - 
    4 m_\pi^2({Q^2}-m_\pi^2)] [{Q^2} - s_+ - m_\pi^2]}
     {{Q^2} + 3 m_\pi^2 - s_+}}
\end{eqnarray}
We perform a least-square fit to the Dalitz plot density as predicted by CHPT, 
using a fit function
\begin{equation}
 \rho_{Q^2}(x,y) :=   \frac{d\Gamma}{d Q^2\,ds_1\,ds_2} = 
   a (x + b + c x^2 + d x^3 + e x^4 + f x (x-1) y^2)
\end{equation}
Note that this is a reasonable ansatz for the $y$ dependence. Firstly,
we must have $\rho_{Q^2}(x,y) = \rho_{Q^2}(x,-y)$ because of Bose symmetry.
Secondly, at $x \to 0$ and at $x\to 1$, $s_-^{max} \to 0$, 
so the dependence on $y^2$ must go to zero at $x=0,1$. 

We choose  to discuss the CHPT predictions for $Q^2 = 0.36\, \mbox{GeV}^2$.
At this value of $Q^2$, the result from the fit is
\begin{eqnarray}
   a & = & (3.0360\pm 0.0095) \times 10^{-14}\, \mbox{GeV}^{-5}
\nonumber \\
   b & = & (5.251 \pm 0.039)\times 10^{-3}
\nonumber \\
   c & = & (0.786 \pm 0.023)
\nonumber \\
   d & = & (0.190 \pm 0.053)
\nonumber \\
   e & = & (0.189 \pm 0.036)
\nonumber \\
   f & = & (0.0645 \pm 0.0026)
\end{eqnarray}
This
fit never deviates from the CHPT prediction by more than $2 \%$, with 
an average deviation of less than $1\%$.
It is seen that the $y^2$ dependence is very small. In fact, taking into
account that the coefficient $x(x-1)$, which multiplies $y^2$,
is less or equal to $1/4$, we find that the
leading $y$ dependence does not exceed $2\%$.
We have checked that, as one would expect, 
the $y^2$ dependence is even smaller for smaller $Q^2$.

This Dalitz plot distribution, as predicted by CHPT for $Q^2 \leq 0.36
\, \mbox{GeV}^2$, 
differs strongly from the behavior at high $Q^2$ in the
resonance regime, where the $\rho$ resonances lead to pronounced
structures in $s_1$ and $s_2$ (resonance bands for fixed $s_1$ or
$s_2$) \cite{argus93}.  

\subsection{Comparison with Vector Meson Dominance Models}
\label{compvmd}
In this subsection we will compare with the low energy
behavior of the phenomenological models in \cite{Kue90,Dec92,Dec94}.
The simplest VMD model which one can build for this channel
(see Ref. \cite{Kue90}) is based on
the decay chain $\tau \to a_1 \nu_\tau$, $a_1 \to \rho \pi$, $\rho \to
2 \pi$ and a transverse $a_1$ propagator. 
In this case the amplitude contains only a spin 1 part ($F_S
=0$), and the three pions are only in a $[210]$ partition state
($F_1^{[300]} =0$). The comparison to the data \cite{argus90,opal}
shows that the model works well, which means that the assumptions made are 
reasonable. 

However it is clear that these assumptions need not be strictly true
in the physical reality, so the authors of Refs. \cite{Dec92,Dec94}
have tried to include in the VMD model a nonzero scalar form factor. 
Two possible sources for a non-vanishing $F_S$, are a pseudoscalar
three-pion resonance, the $\pi'$, or the non-transverse component of
the off-shell $a_1$ propagator. 
A model for the  $\pi'$ contribution is given in \cite{Dec92}. 
The numerical impact of the $\pi'$ depends on a parameter $f_{\pi'}$. 
In \cite{Dec92} and in its implementation in TAUOLA \cite{tauola}
a particularly
large value from \cite{Isg} has been chosen, which is probably several
orders of magnitude too large \cite{Hagi}. 
However, we find that even with this high value of $f_{\pi'}$,
the contribution from the $\pi'$ to the scalar form factor at low energies
is much smaller than the predictions from CHPT.
This indicates that --- at least at small energies --- 
there are additional contributions to $F_S$.
The off-shell contribution of the $a_1$ to $F_S$ is discussed in
\cite{Dec94}. A specific model is constructed by matching to the
$O(p^2)$ prediction of CHPT (including the $M_\pi^2 \neq 0$ effects),
and we will compare this model with the CHPT predictions.

On the other hand, the presence of a $[300]$ component in the spin 1
form factor has never been proposed in any of these models. 
With our calculation we can make a detailed analytical comparison
between VMD models and the CHPT amplitude at low energy.
The main conclusion is that even in the region close to threshold the
VMD model works rather well.

First we consider the structure functions which only involve the 
$F_1$. After a proper normalization of the form
factors in the VMD model, which takes into account the CHPT
expressions at $O(p^2)$, the agreement for the spin one spectral function at
low energy looks very good. 
This can be seen in Fig. \ref{figwa}, where we plot $w_A$,
i.e.\ the main contribution to the total decay rate.
Moreover the $[300]$ component of $F_1$ starts
only at one loop, which means that in the chiral expansion this is
algebraically suppressed with respect to the $[210]$ part.

However, the - numerically rather small - structure functions $w_D$ and
$w_E$ are sensitive to the $[300]$ part via interference with the $[210]$
part, as we have seen in the previous subsection.
Comparing the VMD model with the CHPT prediction for these structure functions
in Figs.~\ref{figwd} and \ref{figwe}, we find good agreement for
$2\pi^0 \pi^-$, but 
flat disagreement for $2\pi^- \pi^+$. 
However, at larger $Q^2$, experimental data for $w_D$ and $w_E$ in
$2\pi^- \pi^+$ 
are available, which agree with the VMD prediction \cite{argus90,opal}.
Note that the left-right asymmetry $A_{LR}(Q^2)$ measured by ARGUS 
\cite{argus90} is proportional to $w_E(Q^2)/w_A(Q^2)$ \cite{KM1}, and 
confirms the VMD prediction in sign and magnitude down to $Q^2 = 0.8\, 
\mbox{GeV}^2.$
We conclude that, unless the higher orders in the chiral expansion completely
change the CHPT result, somewhere between threshold and $Q^2 = 0.8
\mbox{GeV}^2$ there must be a zero for both structure functions in the
$2 \pi^- \pi^+$ mode. It would be extremely interesting to verify this
zero experimentally, or, as a minimal option, to verify the existence
of a difference between the two charge modes. 

Next we consider the scalar form factor: Although $F_S$ is nonzero already
at tree level, it is kinematically suppressed:
\begin{eqnarray}
F_1 &=& \frac{i \sqrt{2}}{F_\pi} \left(\frac{2}{3}+O(p^2)\right) \; ,
\nonumber \\ 
F_S &=& \frac{i \sqrt{2}}{F_\pi} \left(\frac{M_\pi^2}{Q^2}
\frac{s_3-M_\pi^2}{M_\pi^2-Q^2}+ O(p^2)\right) \;.
\end{eqnarray}
Though $Q^2, s_3$ and $M_\pi^2$ are all counted as quantities of
order $p^2$, so that the ratios $M_\pi^2/Q^2$ and $s_3/Q^2$ are
algebraically of order one, it is clear that numerically they are
smaller than one. For example at threshold $F_S =i
\sqrt{2}/F_\pi (-1/24)=-1/16 \times  F_1$. 

In Figs. \ref{figwsaP},\ref{figwsa0} and \ref{figwsb}, where we plot the
two structure functions $w_{SA}$ and $w_{SB}$, one can see the
comparison of the VMD model in \cite{Dec94} 
to CHPT at $O(p^4)$.
We find that the VMD model, which by construction reproduces the $O(p^2)$
prediction, does not reproduce well the $O(p^4)$ prediction from CHPT. 
The reason for this discrepancy (which is larger for $w_{SA}$, 
since it contains the scalar form factor squared) lies in the fact that in
the model in \cite{Dec94}, the $[300]$ part of $F_S$ is added as a constant,
without a resonance factor enhancement.

Summarizing our comparison of CHPT and VMD predictions at low energies,
we have found that VMD gives a good description of the dominant
spin-1, $[210]$  contribution.
However, CHPT shows that both $F_S$ and $F_1^{[300]}$, though
small, are not exactly zero. This fact, which is a prediction of CHPT,
has still to be verified experimentally. Our analysis shows that the
best place to look for these parts of the amplitude, is where they
can interfere with the ``big'' components: the presence of $F_S$
should be detected by measuring a nonzero $w_{SB}$, whereas the
presence of $F_1^{[300]}$ should be discovered by measuring a sizable
difference (possibly a sign difference) between the two charge modes
for $w_D$ and $w_E$.

\section{Summary and Conclusions}
%
Chiral perturbation theory (CHPT) provides model-independent predictions
for hadronic matrix elements in the low energy region below
$500 \cdots 600\,\mbox{MeV}$. It does not contain additional
assumptions beyond the fact that the strong interactions are described
by the QCD Lagrangian and that QCD possesses an approximate chiral
symmetry which is spontaneously broken.
We have evaluated tau decays into two pions (and tau-neutrino) to
two loops and decays into three pions to
the one loop level.
The branching ratio into the phase space region with small enough 
invariant hadronic mass for CHPT to be applicable was found to be about
$4 \times 10^{-3}$ for the two pion mode
and about $10^{-5}$ for the three pion mode.
And so the predictions of CHPT for $\tau \to 2 \pi \nu_\tau$
are testable at present machines (LEP and CESR),
while in the case of $\tau \to 3 \pi \nu_\tau$, future facilities
with very high $\tau$ production rates seem to be required (b factories,
$\tau$-charm factory).

In the case of the decay  $\tau \to 2 \pi \nu_\tau$, 
the predictions for the invariant mass distributions can
be tested at present experiments, 
and the spectrum can be used to extract the pion
charge radius $\langle r^2 \rangle_V^\pi$.

As for the decay $\tau \to 3 \pi \nu_\tau$, a detailed comparison to
the CHPT predictions for the form factors near threshold 
requires very high statistics,  mainly because of the
phase space suppression. 
For this reason we have tried to identify a few spots where the
consequences of the approximate chiral symmetry of QCD can be tested
experimentally with a reasonable statistics. 
These are:
\begin{enumerate}
\item
the very close similarity of $w_A$ and $w_C$ near threshold, which
seems to extend well beyond the very low energy region;
\item
a zero and change of sign
in $w_D$ and $w_E$ between threshold and $Q^2 \sim 0.8
\mbox{GeV}^2$ in the channel $ 2 \pi^- \pi^+$, or at least a
difference between the two charge modes near threshold. This would be
the first evidence for the presence of the $[300]$ partition state in
this decay channel;
\item
the presence of a scalar form factor as predicted by CHPT, to be
detected by measuring $w_{SB}$.
\end{enumerate}

Comparing our results to predictions from vector meson dominance
models, we find overall a reasonable agreement in the
low energy region. The structure function
$w_A$, which dominates the decay rate, is described well by VMD models.
We have found, however, some interesting discrepancies in certain 
(numerically rather small) structure functions. 
These discrepancies are related to the [300] partition state and to the
scalar form factor, which both are missing or underestimated in vector
meson dominance models.
%
\section*{Acknowledgments}
%
We would like to thank J.H.\ K\"uhn for bringing this subject to
our attention. We also thank him and J. Gasser for very helpful
discussions and for carefully reading the manuscript.
Further thanks are due to Erwin Mirkes, for supplying us with the
code of his tau Monte-Carlo program and for his helpful comments.

This work is supported in part by the National Science
Foundation (Grant \#PHY-9218167), by HCM, EEC-Contract No.
CHRX-CT920026 (EURODA$\Phi$NE), by BBW, Contract No. 93.0341, by the
Deutsche Forschungsgemeinschaft, and by Schweizerischer Nationalfonds.

R.U. thanks the members of the Institute for Theoretical Physics in 
Berne for their kind hospitality during the final stage of this work.
M.F. thanks the members of the Theoretical Physics Group at Harvard
University for their kind hospitality.

\appendix
\section{Appendix}
%
\subsection{General conventions}
\label{app1}
Consider the decay of a tau into $n$ pions,
\begin{equation}
   \tau(p,s) \to \nu_\tau(q,s') \pi_1(p_1) \pi_2(p_2) \cdots
   \pi_n(p_n)
\end{equation}
Here $s$ and $s'$ denote the polarization four-vectors of the tau and the
neutrino, respectively.
We define the total hadronic momentum $Q$ by
\begin{equation}
   Q = p_1 + p_2 + \cdots + p_n
\end{equation}
and in the case of three pions, we will use Dalitz plot invariants
$s_1$, $s_2$ and $s_3$ defined by
\begin{equation}
    s_1 = (p_2 + p_3)^2
\end{equation}
and cyclic permutations.

The differential decay rate $d\Gamma_n$ is given by
\begin{equation}
  d\Gamma_n = \frac{1}{2 M_\tau} |{\cal M}|^2 d\Phi_n
\end{equation}
where the phase space element $d\Phi_n$ is
\begin{equation}
   d \Phi_n = 
   (2 \pi)^4 \delta( Q + q - p)
   \frac{d^3 q}{(2 \pi)^3 2 E_\nu}
   \prod_{k=1}^{n} \frac{d^3 p_k}{(2 \pi)^3 2 E_k} 
\end{equation}
and
\begin{eqnarray} \label{ccb}
   {\cal M} & = & \frac{G_F}{\sqrt{2}} \cos \theta_C L_\mu H^\mu
\nonumber \\
   L_\mu & = & \overline{u}_\nu(q,s') \gamma_\mu \gamma_- u_\tau(p,s)
\nonumber \\
   H_\mu & = & \langle \pi_1(p_1) \pi_2(p_2) \dots \pi_n(p_n) 
     | V_\mu(0) - A_\mu(0)| 0 \rangle 
\end{eqnarray}

\subsection{Decomposition of the form factors in terms of partition
states}
\label{partition}
As we discussed in Sect. \ref{secpartitions}, we have three possible
partition states, two belonging to the class $[210]$ (which we will
indicate as $[210]_a$ and $[210]_b$) and one belonging
to the class $[300]$. The matrix elements of these states are easily
constructed by performing the appropriate symmetrizations and
antisymmetrizations with respect to momenta exchanges, as described in
Sect. \ref{secpartitions}. The result can be expressed in terms of the
three form factors $F, \; G$ and $H$, and reads as follows:
\begin{eqnarray}
F^{[210]_a} &=& \frac{2}{3} \left[ F_{12} - G_{23} + H_{23} \right] 
\nonumber \\ 
G^{[210]_a} &=& \frac{2}{3} \left[ G_{12} -\frac{1}{2}\left( F_{23} +
G_{23}+ H_{23} \right) \right] 
\nonumber \\ 
H^{[210]_a} &=& \frac{2}{3} \left[ H_{12} -\frac{1}{2}\left( F_{23} -
G_{23} - H_{23} \right) \right] \; ,
\end{eqnarray}
\begin{eqnarray}
F^{[210]_b} &=& \frac{1}{3} \left[ - G_{13} + G_{23} + H_{13} - H_{23} \right]
\nonumber \\ 
G^{[210]_b} &=& \frac{1}{6} \left[ - F_{13} + F_{23}
-G_{13}+G_{23}-H_{13} + H_{23}  \right]
\nonumber \\ 
H^{[210]_b} &=& \frac{1}{6} \left[ F_{13} + F_{23}
-G_{13}-G_{23}-H_{13} - H_{23}  \right] \; ,
\end{eqnarray}
\begin{eqnarray}
F^{[300]} &=& \frac{1}{3} \left[ F_{12} +G_{13}+ G_{23} -H_{13}-
H_{23} \right] 
\nonumber \\ 
G^{[300]} &=& \frac{1}{3} \left[ G_{12} +\frac{1}{2}\left( F_{13}+F_{23} +
G_{13}+G_{23} + H_{13}+ H_{23} \right) \right]
\nonumber \\ 
H^{[300]} &=& \frac{1}{3} \left[ H_{12} + \frac{1}{2}\left(-F_{13}+
F_{23} +G_{13} - G_{23} +H_{13} - H_{23} \right) \right] \; ,
\end{eqnarray}
where $X_{ij}=X(s_i,s_j,s_k)$, $X=F,G,H$ and $k\neq i \neq j$. To
reconstruct the form factors in the two charge modes one has to use
the relations: 
\begin{eqnarray}
X_{12}&=& X^{[300]}+X^{[210]_a}+X^{[210]_b} \nonumber \\
X^{(+)}_{12}&=& 2 X^{[300]}-X^{[210]_a}-X^{[210]_b} \; .
\end{eqnarray}

\subsection{Structure functions for the three pion final state}
\label{app2}
In this section we will briefly review the formalism of structure 
functions for the decay of the tau into three pions, and display
formulae relevant for the present paper. For more details, the
reader is referred to \cite{KM1,KM2}.

The decays are most easily analyzed in the hadronic rest frame
\begin{equation}
   \vec{0} = \vec{Q} = \vec{p_1} + \vec{p_2} + \vec{p_3}
\end{equation}
The orientation of the hadronic system is characterized by three Euler
angles $\alpha$, $\beta$, $\gamma$, as introduced in \cite{KM1,KM2}.
They can be defined by
($ 0 \leq \alpha, \gamma < 2 \pi$; $0 \leq \beta < \pi$)
\begin{eqnarray}
   \cos \alpha = \frac{(\vec{n}_L \times \vec{n}_\tau) \cdot
   (\vec{n}_L \times \vec{n}_\perp)}
   {|\vec{n}_L \times \vec{n}_\tau| |\vec{n}_L \times \vec{n}_\perp|}
& & 
   \sin \alpha = - \frac{\vec{n}_\tau \cdot
   (\vec{n}_L \times \vec{n}_\perp)  }
   {|\vec{n}_L \times \vec{n}_\tau| |\vec{n}_L \times \vec{n}_\perp|}
\nonumber \\
\nonumber \\
   \cos \gamma = - \frac{\vec{n}_L \cdot \vec{n}_3}
   { \vec{n}_L \times \vec{n}_\perp}
& &
   \sin \gamma = \frac{(\vec{n}_L \times \vec{n}_\perp) \cdot \vec{n}_3}
   { \vec{n}_L \times \vec{n}_\perp}
\nonumber \\
\nonumber \\
    \cos \beta =  \vec{n}_L \cdot \vec{n}_\perp       
\end{eqnarray}
where $\vec{n}_L$ is the direction of the laboratory in the hadronic rest
frame, $\vec{n}_\perp = (\vec{p}_1 \times \vec{p}_2)/
(|\vec{p}_1 \times \vec{p}_2|)$ is the normal to the pion plane
(here we assign the momenta according to $|\vec{p_2}| >  |\vec{p_1}|$),
$\vec{n}_\tau$ is the direction of flight of the $\tau$ in the hadronic
rest frame and $\vec{n}_3 = \vec{p}_3 / |\vec{p}_3|$.

{}From these definition it is obvious that $\beta$ and $\gamma$ are 
observable even if the tau rest frame cannot be reconstructed, 
whereas $\alpha$ does require this knowledge. 
$\alpha$ could be measured at a tau-charm factory
where the tau pairs are produced almost at rest and therefore the
tau rest frame is known, or if the tau direction can be measured with
the help of vertex detectors \cite{Kue93}.

The Euler angles also serve to parameterize the phase space
\begin{equation}
   d PS^{(4)} = \frac{1}{(2 \pi)^5} \frac{1}{64}
   \frac{M_\tau^2 - Q^2}{M_\tau^2} \frac{dQ^2}{Q^2} ds_1\,ds_2
   \frac{d \alpha}{2 \pi} \frac{d \gamma}{2 \pi}
   \frac{d \cos \beta}{2} \frac{d \cos \theta}{2}
\end{equation}
where $\theta$ is the angle between the direction of flight of the $\tau$
in the laboratory frame and the direction of the pions as seen in the
$\tau$ rest frame, and $\cos \theta$ can be calculated from the
energy $E_h$ of the pion system with respect to to the laboratory frame
and beam energy $E_{beam}$
\begin{equation}
   \cos \theta = \frac{ 2 E_h/E_b M_\tau^2 - M_\tau^2 - Q^2}
   {(M_\tau^2 - Q^2)
   \sqrt{1 - M_\tau^2/ E_{beam}^2}}
\end{equation}
The contraction of the leptonic and hadronic tensors can be expanded
in a sum 
\begin{equation}
    L_{\mu\nu} H^{\mu\nu} = \sum_{X} L_X W_X
\end{equation}
In general, $H^{\mu\nu}$ can be characterized by 16 independent real 
functions. In our case of a three pion final states, there are restrictions
due to $G$ parity and Bose symmetry, which leave 9 independent functions.
In a convenient basis they are given by \cite{KM2}
\begin{eqnarray}  \hspace{3mm}
W_{A}  &=&   \hspace{3mm}(x_{1}^{2}+x_{3}^{2})\,|F_{1}|^{2}
           +(x_{2}^{2}+x_{3}^{2})\,|F_{2}|^{2}
           +2(x_{1}x_{2}-x_{3}^{2})\,\mbox{Re}\left(F_{1}F^{\ast}_{2}\right)
                                   \nonumber \\[1mm]
W_{B}  &=& \hspace{3mm} x_{4}^{2}|F_{3}|^{2}
                                   \nonumber \\[1mm]
W_{C}  &=&  \hspace{3mm} (x_{1}^{2}-x_{3}^{2})\,|F_{1}|^{2}
           +(x_{2}^{2}-x_{3}^{2})\,|F_{2}|^{2}
           +2(x_{1}x_{2}+x_{3}^{2})\,\mbox{Re}\left(F_{1}F^{\ast}_{2}\right)
                                   \nonumber \\[1mm]
W_{D}  &=&  \hspace{3mm}2\left[ x_{1}x_{3}\,|F_{1}|^{2}
           -x_{2}x_{3}\,|F_{2}|^{2}
           +x_{3}(x_{2}-x_{1})\,\mbox{Re}\left(F_{1}F^{\ast}_{2}\right)\right]
                                   \nonumber \\[1mm]
W_{E}  &=& -2x_{3}(x_{1}+x_{2})\,\mbox{Im}\left(F_{1}
                    F^{\ast}_{2} \right)\nonumber \\[1mm]
W_{F}  &=&  \hspace{3mm}
          2x_{4}\left[x_{1}\,\mbox{Im}\left(F_{1}F^{\ast}_{3}\right)
                     + x_{2}\,\mbox{Im}\left(F_{2}F^{\ast}_{3}\right)\right]
                                   \nonumber \\[1mm]
W_{G}  &=&- 2x_{4}\left[x_{1}\,\mbox{Re}\left(F_{1}F^{\ast}_{3}\right)
                     + x_{2}\,\mbox{Re}\left(F_{2}F^{\ast}_{3}\right)\right]]
                                   \nonumber \\[1mm]
W_{H}  &=& \hspace{3mm}
      2x_{3}x_{4}\left[\,\mbox{Im}\left(F_{1}F^{\ast}_{3}\right)
                     -\,\mbox{Im}\left(F_{2}F^{\ast}_{3}\right)\right]
                                   \label{wi} \\[1mm]
W_{I}  &=&- 2x_{3}x_{4}\left[\,\mbox{Re}\left(F_{1}F^{\ast}_{3}\right)
                     -\,\mbox{Re}\left(F_{2}F^{\ast}_{3}\right)\right]
                                   \nonumber \\[1mm]
W_{SA} &=& \hspace{3mm} Q^{2}\,|F_{4}|^{2}\nonumber\\[1mm]
W_{SB} &=&  \hspace{3mm}2\sqrt{Q^{2}}\left[
            x_{1}\,\mbox{Re}\left(F_{1}F^{\ast}_{4}\right)
           +x_{2}\,\mbox{Re}\left(F_{2}F^{\ast}_{4}\right)
             \right]\nonumber\\[1mm]
W_{SC} &=&- 2\sqrt{Q^{2}}\left[
            x_{1}\,\mbox{Im}\left(F_{1}F^{\ast}_{4}\right)
           +x_{2}\,\mbox{Im}\left(F_{2}F^{\ast}_{4}\right)
             \right]\nonumber \\[1mm]
W_{SD} &=& \hspace{3mm} 2\sqrt{Q^{2}}x_{3}\left[
            \,\mbox{Re}\left(F_{1}F^{\ast}_{4}\right)
           -\,\mbox{Re}\left(F_{2}F^{\ast}_{4}\right)
             \right]\nonumber \\[1mm]\nonumber
W_{SE} &=& -2\sqrt{Q^{2}}x_{3}\left[
            \,\mbox{Im}\left(F_{1}F^{\ast}_{4}\right)
           -\,\mbox{Im}\left(F_{2}F^{\ast}_{4}\right)
             \right]\\[1mm]\nonumber
W_{SF} &=&- 2\sqrt{Q^{2}}x_{4}\,\mbox{Im}\left(F_{3}F^{\ast}_{4}\right)
                             \\[1mm]\nonumber
W_{SG} &=& -2\sqrt{Q^{2}}x_{4}\,\mbox{Re}\left(F_{3}F^{\ast}_{4}\right)
                             \nonumber
\end{eqnarray}
The variables $x_i$ are defined by
   \begin{eqnarray}
x_{1}& = & q_{1}^{x}-q_{3}^{x}  \nonumber \\
x_{2} & = & q_{2}^{x}-q_{3}^{x} \nonumber \\
x_{3} & = & q_{1}^{y}=-q_{2}^{y} \nonumber \\
x_{4} & = & \sqrt{Q^{2}}x_{3}q_{3}^{x}
   \end{eqnarray}
where $q_i^{x}$ ($q_i^{y}$) denotes
the $x$ ($y$) component of the momentum of
meson $i$ in the hadronic rest frame.
They can easily be expressed in terms of $s_1$, $s_2$ and $s_3$
\cite{KM1,KM2}.

The hadronic structure functions $W_X$ depend only on $s_1$, $s_2$ and
$Q^2$. The corresponding leptonic $L_X$ depend 
on the Euler angles $\alpha$, $\beta$, $\gamma$ and on $E_h$. The
relevant formulae can be found in \cite{KM2}.

The structure functions can be measured by observing angular distributions 
in the Euler angles 
$\beta$, $\gamma$ and, if the tau rest frame is known, $\alpha$, 
and taking moments $\langle m\rangle $ with respect to 
products of trigonometric functions
\begin{equation}
   \langle m\rangle  := \frac{3}{2 (M_\tau^2 - Q^2)} \int 
   L_{\mu\nu} H^{\mu \nu} \, m \,\frac{d \cos \beta}{2} 
   \frac{d \gamma}{2 \pi}
\end{equation}
As shown in \cite{KM2}, measuring
suitable moments involving $\beta$ and $\gamma$
only (and, in fact, an energy ordering $\mbox{sign}(s_1 -s_2)$ in some
cases to avoid vanishing due to Bose symmetry) allows to extract
all the individual structure functions except for $W_{SC}$ and $W_{SE}$.
If the tau rest frame and hence $\alpha$ is known additionally, $W_{SC}$
and $W_{SE}$ can be measured, too.


\end{document}